\documentclass[twocolumn,groupedaddress,aps,pra,showpacs,10pt]{revtex4-2}
\usepackage[colorlinks,linkcolor=blue,anchorcolor=blue,citecolor=blue,bookmarksnumbered]{hyperref}
\usepackage{graphicx,epsfig,float,epstopdf}
\usepackage{amssymb,amsmath,subeqnarray,mathrsfs,amsfonts,bm,makecell,cases,color,extpfeil,lipsum}
\usepackage[normalem]{ulem}
\usepackage{graphicx}
\usepackage{dcolumn}
\usepackage{bm}
\allowdisplaybreaks

\begin{document}

\title{Giant and sign-tunable spin Hall effect of light in a nonlocal nonlinear Rydberg medium}
\author{Ying Sun$^{1}$}
\author{Lu Qin$^{1,2,3}$*} 
\author{Zeyun Shi$^{4}$} 
\author{Yingying Zhang$^{1}$}
\author{Zunlue Zhu$^{1}$*}
\author{Weibin Li$^{2,3}$}
\author{Xingdong Zhao$^{1}$*}%

\affiliation{
$^{1}$\mbox{School of Physics, Henan Normal University, Xinxiang 453007, China}\\
$^{2}$School of Physics and Astronomy, University of Nottingham, Nottingham NG7 2RD, United Kingdom\looseness=-1\\
$^{3}$Centre for the Mathematics and Theoretical Physics of Quantum Non-equilibrium Systems, University of Nottingham, Nottingham NG7 2RD, United Kingdom\looseness=-1\\
$^4$\mbox{School of Artificial Intelligence, Hubei University of Automotive Technology,} Shiyan 442000, China\looseness=-1\\
}

\thanks{qinlu@htu.edu.cn\\
zl-zhu@htu.edu.cn\\
phyzhxd@gmail.com}

\date{\today}

\begin{abstract}
We demonstrate that long-range Rydberg atom interactions provide a tunable mechanism for generating the giant spin Hall effect of light (SHEL) in a cold atomic medium.
For a coherently driven $^{87}\mathrm{Rb}$ medium with Rydberg $S$ states,
 we derive the linear, local, and nonlocal Kerr susceptibilities for the two orthogonal probe polarization components and evaluate the resulting spatial and angular SHEL shifts. The nonlocal third-order susceptibility  reshapes the complex birefringence of the medium, and  drastically enhances the $y$-polarized SHEL shift by about one to two orders of magnitude. Furthermore, it can reverse the direction of both the spatial and angular shifts relative to the  predictions of linear response. The effect is controllably tuned by the laser detuning, the quantization-axis orientation, the atomic density, and the principal quantum number of the Rydberg state.
 In the $^{87}$Rb medium, we find that the spatial and angular shifts reverse sign when the principal quantum number of the Rydberg state are around $n=40$ and 60, respectively.
At high-lying excited Rydberg states, the spatial shift can be enhanced to several tens of micrometres—a magnitude that should be directly observable in experiment.
These results demonstrate that Rydberg-mediated nonlocal optical nonlinearity is an efficient mechanism for spin-dependent light transport and for future photonic sensing applications.


\end{abstract}

\maketitle

\section{Introduction}

The spin--orbit interaction (SOI) of light, which couples the spin angular momentum associated with polarization to the orbital degrees of freedom of light, is the foundation for a broad range of spin-dependent optical phenomena~\cite{Bliokh2015SpinOrbit,marrucci2006optical,Petersen2014,Qin_2026}. A representative manifestation is the spin Hall effect of light (SHEL), in which the right- and left-circularly polarized components of a light beam undergo opposite transverse shifts during propagation, reflection, or transmission~\cite{Onoda2004Hall,Kim2023,Imbert1972,korger2014observation}. The SHEL provides a photonic analogue of the electronic spin Hall effect and has attracted considerable interest because of its fundamental connection to geometric phases and angular-momentum conservation~\cite{Bliokh2004Topological,Bliokh2006Conservation,ling2017recent,Sheng2023PSHE}, as well as its potential applications in precision metrology~\cite{Zhou2012Graphene,Zhou2012Nanometal,Chen2020}, optical sensing\cite{WangB2020,Neugebauer2019,Wang2020APL}, image processing\cite{Zhu2019Differentiation,Tang2024,HeY2025,HeS2020}, and spin-controlled photonic devices~\cite{Yin2013Photonic,HeA2021,Kapitanova2014}. 
However, the spin-dependent transverse shifts generated at conventional optical interfaces are typically far smaller than the optical wavelength, making direct observation challenging and often necessitating the use of weak-value amplification~\cite{Hosten2008,Gorodetski2012PRL} or other highly sensitive detection techniques~\cite{Luo2011Layered,Slobozhanyuk2016Topological}. Developing efficient mechanisms for substantially enhancing the SHEL remains an important and actively pursued objective.

Coherently driven atomic media provide an attractive platform for generating and controlling SHEL because their dispersive and absorptive properties can be manipulated through the frequencies, intensities, and polarizations of external optical fields~\cite{Waseem2024,Shah2025,Liu2025RydbergPSHE}. In multilevel atomic systems, the polarization selection rules associated with different Zeeman transitions allow an initially isotropic atomic ensemble to acquire a controllable optical anisotropy~\cite{Patnaik2000,Datsyuk2008,Wielandy1998}. 
In particular, optical pumping can prepare the atomic population in selected magnetic sublevels,
whereby probe fields polarized parallel and perpendicular to the quantization axis to couple to different $\pi$ and $\sigma^{\pm}$ transitions, resulting in distinct susceptibilities~\cite{Boozer2007,Chen2002Quantization,Gerasimov2011,Xue2019Rydberg}.
The resulting polarization-dependent refractive indices, together with the spin-redirection geometric phase of the angular-spectrum components of a finite-width beam, generate an effective coupling between the photon spin and transverse momentum. It has been shown that this mechanism can produce spatial and angular SHEL shifts considerably larger than those arising at ordinary dielectric interfaces, while allowing their magnitude and direction to be controlled through the optical detuning and orientation of the atomic quantization axis~\cite{Wu2020spin,Wu2019arxiv}.

The polarization sensitivity of atomic optical responses stems from the quantization of angular momentum
and the associated dipole-selection rules. Recent studies on Rydberg-atomic polarimetry have further demonstrated that rotating the polarization of an applied electromagnetic field can strongly reshape electromagnetically induced transparency (EIT) spectra by activating different $\pi$ and $\sigma^{\pm}$ excitation pathways~\cite{Cloutman2026RydbergEIT}. 
These results highlight the ability of multilevel Rydberg atoms to convert the orientation and polarization of external fields into pronounced changes in the optical response, thereby providing a powerful basis for controlling polarization-dependent light propagation.

Previous investigations of the SHEL in coherently prepared atomic media have focused predominantly on linear and local nonlinear optical responses~\cite{Wu2020spin,Abbas2025Tripod,Khan2025EIT,Wan2020}. Although field-induced linear anisotropy offers substantial tunability, the attainable transverse shifts are ultimately constrained by the magnitude of the polarization-dependent susceptibility and by the trade-off between strong dispersion and resonant absorption. Optical nonlinearities introduce an additional degree of freedom for reshaping the complex refractive indices and may therefore provide a more powerful mechanism for enhancing and controlling spin-dependent beam shifts. In conventional atomic media, however, the Kerr nonlinearity is generally weak, while increasing the atom--light interaction near resonance is usually accompanied by strong absorption~\cite{Boyd2008}. EIT can partly overcome this limitation by suppressing resonant absorption through quantum interference while retaining an appreciable nonlinear response~\cite{Fleischhauer2005}; however, the local Kerr nonlinearity of non-interacting atoms often remains too weak to produce a substantial modification of the beam shifts.

Highly excited Rydberg atoms offer a promising route to overcome this limitation. Their large electric dipole moments and long-range Rydberg--Rydberg interactions give rise to strong interatomic correlations and the Rydberg-blockade effect, through which the excitation of one atom modifies the optical response of surrounding atoms over a mesoscopic distance~\cite{Saffman2010,Gallagher2008}. When combined with EIT, these correlations generate a giant nonlocal Kerr nonlinearity that can greatly exceed the corresponding local nonlinear response even at relatively low atomic densities~\cite{Zhang2025Microwave,Bai2016OE,Pritchard2013in}. The importance of this mechanism for optical beam shifts has been demonstrated in studies of the Goos--H\"anchen effect in Rydberg atomic gases~\cite{WuJH2022PRA}.
These findings naturally raise the question of whether Rydberg-mediated nonlocal optical nonlinearities can also be exploited to enhance and control spin-dependent transverse shifts.

In this work we identify the nonlocal Kerr susceptibility, arising from long-range Rydberg-Rydberg interactions, as the dominant nonlinear
mechanism shaping the SHEL in a cold Rydberg atomic medium.  A linearly polarized control field defines the atomic quantization axis and induces polarization-dependent optical responses for the two orthogonal components of a probe beam. By deriving the corresponding linear, local nonlinear, and nonlocal nonlinear susceptibilities, we show that the nonlocal third-order contribution can dominate the nonlinear modification of both the dispersive and absorptive anisotropies. The resulting spatial and angular transverse shifts are highly sensitive to the two-photon detuning and to the angle between the quantization axis and the probe propagation direction. In particular, the shifts of the $y$-polarized component can exceed those of the $x$-polarized component by factors of several tens and, under appropriate conditions, by up to two orders of magnitude. Moreover, the nonlocal nonlinear response not only enhances the magnitude of the SHEL but can also reverse the signs of the spatial and angular shifts.
We further demonstrate that the SHEL can be efficiently tailored by varying experimentally accessible parameters, including the atomic density and the principal quantum number of the Rydberg state. Due to both parameters strongly influence the Rydberg-mediated nonlocal susceptibility, they provide additional means of controlling the magnitude and direction of the spin-dependent transverse transport. Our results establish nonlocal optical nonlinearity in a Rydberg atomic medium as an efficient and highly tunable mechanism for manipulating the SHEL, and reveal that the Rydberg-mediated nonlocal Kerr nonlinearity enhances and controls the SHEL. Our study thus extends previous studies based primarily on linear atomic anisotropy and opening new avenues for precision optical sensing and spin-dependent photonic control.

This paper is organized as follows. Section~\ref{Sec:II} introduces the physical model under study and the spin Hall effect of light (SHEL). Section~\ref{sec:atomic susceptibilities} provides the tunable optical susceptibility of the nonlocal Rydberg atomic medium. Section~\ref{sec:IV} presents the results of the tunable and enhanced SHEL. Section~\ref{sec:V} summarizes the main results of this paper.

\section{Spin Hall effect of light in atomic gases}\label{Sec:II}

\subsection{Physical model}
\begin{figure}
\includegraphics[scale=0.25]{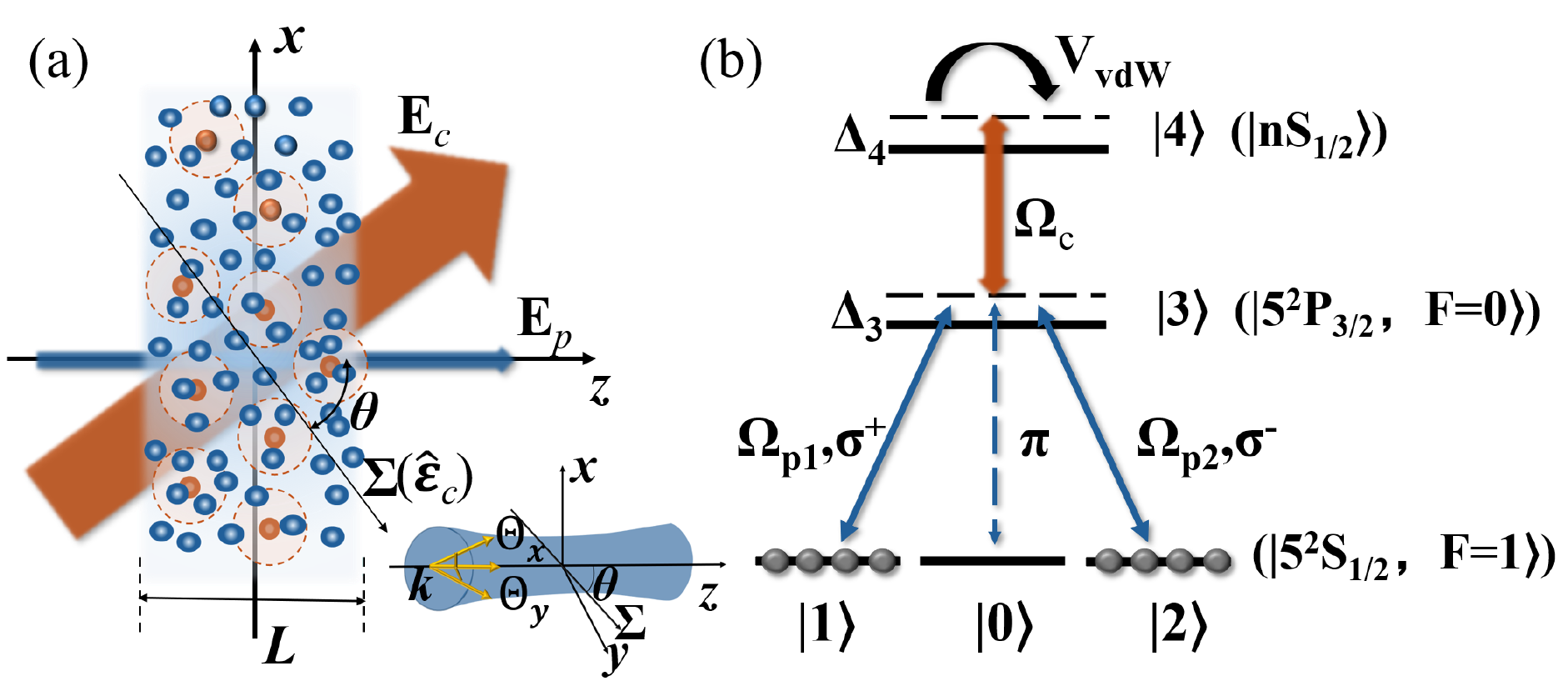}
\caption{
Schematic of the model. (a) A weak probe beam (blue) and a strong control beam (red) propagate through an atomic medium of length $L$. The red and blue solid circles denote atoms excited to the Rydberg state and those remaining unexcited, respectively, while the red dashed circles indicate the boundaries of the corresponding Rydberg-blockade spheres.
Here, $\boldsymbol{\Sigma}$ denotes the quantization axis, and $\theta$ is the angle between the quantization axis and the $z$ axis. (b) Relevant energy levels of the $^{87}\mathrm{Rb}$ D$_2$ line. The solid and dashed blue lines represent the $\sigma^{\pm}$ and $\pi$ transitions driven by the probe field, respectively, whereas the solid red line represents the $\pi$ transition driven by the linearly polarized control field. Here, $\Delta_j$ denotes the corresponding detuning, and $\Gamma_{jl}$ represents the rate of spontaneous decay or incoherent population transfer from state $|j\rangle$ to state $|l\rangle$. The interaction between two atoms excited to the Rydberg state $|4\rangle$ at different positions is described by the van der Waals potential $V_{\mathrm{vdW}}$. The atoms are assumed to be initially prepared in the outermost Zeeman sublevels with $m_F=\pm1$.
 }\label{fig:model}
\end{figure}

We consider a laser-cooled atomic gas of $^{87}\mathrm{Rb}$ as a realistic and experimentally feasible example, as illustrated in Fig.~\ref{fig:model}(a). The linearly polarized control field, $\mathbf{E}_c$, has a polarization vector $\hat{\boldsymbol{\epsilon}}_c$ that lies in the $x$-$z$ plane at an angle $\theta$ relative to the $z$ axis, such that
$\hat{\boldsymbol{\epsilon}}_c
=-\hat{\mathbf{x}}\sin\theta
+\hat{\mathbf{z}}\cos\theta$.
Here $\hat{\mathbf{x}}$, $\hat{\mathbf{y}}$, and $\hat{\mathbf{z}}$ are the basis vectors along the $x$, $y$, and $z$ axes, respectively.
We choose the polarization direction of the control field as the quantization axis, i.e.,
$\boldsymbol{\Sigma}=\hat{\boldsymbol{\epsilon}}_c$. The control field is therefore $\pi$ polarized with respect to the quantization axis and drives the transition
$|5\,^{2}P_{3/2},F=0\rangle
\leftrightarrow
|n\,S_{1/2}\rangle$, as shown in Fig.~\ref{fig:model}(b). 
The probe field drives the D2 transition
 $|5\,^{2}S_{1/2},F=1\rangle
\leftrightarrow
|5\,^{2}P_{3/2},F=0\rangle$.
The corresponding excitation pathways depend on the orientation of the probe polarization relative to the quantization axis. We assume that the atoms are initially prepared in the two outermost Zeeman sublevels $m_F=\pm1$, denoted by $|1\rangle$ and $|2\rangle$, respectively. 
The control field defines the quantization axis, breaking the isotropic optical response of the atomic ensemble because different probe polarization components obey different dipole selection rules.




For a probe field polarized along the $\hat{\mathbf{y}}$ direction, its polarization is always perpendicular to quantization axis $\boldsymbol{\Sigma}$, because the quantization axis lies in the $x$--$z$ plane. The $y$-polarized field can therefore be decomposed into $\sigma^+$ and $\sigma^-$ components and drives the transitions from the initially populated $m_F=\pm1$ ground states to the $m_F=0$ excited state. The effective refractive index experienced by the $y$-polarized probe field is consequently given by
\begin{equation}
n_y
=
\sqrt{\varepsilon_\sigma},
\label{eq:ny}
\end{equation}
where
$\varepsilon_\sigma=1+\chi$
is the relative permittivity associated with the $\sigma^\pm$ transitions, and $\chi$ denotes the corresponding optical susceptibility.

For a probe field polarized along the $\hat{\mathbf{x}}$ direction, the polarization generally contains both components parallel and perpendicular to the quantization axis. The perpendicular component drives the allowed $\sigma^\pm$ transitions. By contrast, the parallel component is $\pi$ polarized and obeys $\Delta m_F=0$. Because the atoms initially occupy only the $m_F=\pm1$ ground states,
 no corresponding $\pi$ transition is available. The parallel component therefore does not couple to the atoms, and its relative permittivity is $\varepsilon_\pi=1$~\cite{Wu2020spin,Wu2019arxiv}.
The effective refractive index experienced by the $x$-polarized probe field is then
\begin{equation}
n_x
=
\sqrt{
\frac{\varepsilon_\sigma}
{\varepsilon_\sigma\sin^2\theta+\cos^2\theta}
}.
\label{eq:nx}
\end{equation}

Consequently, the two orthogonal probe-polarization components generally experience different refractive indices, $n_x\neq n_y$. The atomic gas thus behaves as an effective anisotropic medium whose optical axis is determined by the control field polarization. Both the magnitude and orientation of this anisotropy can be tuned through experimentally accessible control parameters, providing a versatile platform for manipulating the SHEL.

\subsection{Spin Hall effect of light}

Based on the above discussion, we now investigate the spin-dependent transverse shifts of a linearly polarized paraxial Gaussian probe beam. The incident field is written as $\mathbf{E}_p(x,y,z)
\propto\hat{\boldsymbol{\epsilon}}_p u(x,y)\exp(ikz)$, with $u(x,y)={1}{/w_0}\exp\left[-(x^2+y^2)/{w_0^2}\right]$ the transverse field envelope, $w_0$  the beam waist, $\hat{\boldsymbol{\epsilon}}_p$  the polarization vector, and $k$ is the wave number. In the angular-spectrum representation, the Gaussian beam is treated as a continuous superposition of plane-wave components,
$\mathbf{E}_p(\mathbf{r})
\propto
\iint
\widetilde{\mathbf{E}}_p(\Theta_x,\Theta_y)
e^{i\mathbf{k}\cdot\mathbf{r}}
\,d\Theta_x\,d\Theta_y$.
The angular spectrum of the transverse electric field can be expressed as
\begin{equation}
\widetilde{\mathbf{E}}_p(\Theta_x,\Theta_y)
\propto
\begin{pmatrix}
\xi\\
\eta
\end{pmatrix}
\widetilde{u}(\Theta_x,\Theta_y),
\label{eq:incident_angular_spectrum}
\end{equation}
where $(\xi,\eta)^{T}$ is the normalized Jones vector of the incident polarization in the $(x,y)$ basis, satisfying
$|\xi|^2+|\eta|^2=1$.
$\Theta_j={k_j}/{k}$ ($j=x,\,y$)
denotes the small angular deviation of each plane-wave component from the central propagation direction, with
$\Theta_x^2+\Theta_y^2\ll1$.
The Fourier transform of the Gaussian envelope is $\widetilde{u}(\Theta_x,\Theta_y)=w_0/2\exp\left[-(\Theta_x^2+\Theta_y^2){(k w_0)^2}/{4}\right]$,
which describes the amplitude distribution of the probe beam in momentum space.

We next consider the transmission of the probe beam through the anisotropic atomic medium. For each plane-wave component, the incident and transmitted angular spectra are related through
\begin{equation}
\widetilde{\mathbf{E}}'_p(\Theta_x,\Theta_y)
=
\widetilde{M}(\Theta_x,\Theta_y)
\widetilde{\mathbf{E}}_p(\Theta_x,\Theta_y),
\notag
\end{equation}
where, within the paraxial and first-order approximations (see Appendix~\ref{AP:SHEL}), the modified Jones matrix is given by~\cite{bliokh2016spin,Bliokh2019}
\begin{equation}
\widetilde{M}(\Theta_x,\Theta_y)
\simeq
\begin{pmatrix}
e^{i\phi_x(\theta)}
\left[
1+i\Theta_x
\dfrac{\partial\phi_x}{\partial\theta}
\right]
&
e^{i\phi_x(\theta)}\Theta_yY_x
\\[10pt]
-e^{i\phi_y}\Theta_yY_y
&
e^{i\phi_y}
\end{pmatrix}.
\label{eq:Mtheta_total}
\end{equation}
Here
$Y_{x,y}
=
\left(1-e^{\pm i\Phi}\right)\cot\theta$.
The complex phase difference accumulated by the two orthogonal polarization components is denoted by
$\Phi\equiv\phi_y-\phi_x$. The individual phases are
\begin{equation}
\phi_x=n_xkL,
\qquad
\phi_y=n_ykL,\notag
\end{equation}
with $L$ denoting the length of the atomic medium. 
The real part of $\Phi$, $\Phi'=\operatorname{Re}(\Phi)$,
characterizes the dispersive anisotropy, whereas the imaginary part
$\Phi''=\operatorname{Im}(\Phi)$
characterizes the absorptive anisotropy of the atomic medium.

For an $x$-polarized incident probe beam,
$\hat{\boldsymbol{\epsilon}}_p=
\hat{\mathbf{x}}=
({\hat{\mathbf{e}}_{+}
+\hat{\mathbf{e}}_{-}})/{\sqrt{2}}$,
where
$\hat{\mathbf{e}}_{\pm}=
(\hat{\mathbf{x}}\pm i\hat{\mathbf{y}})/\sqrt{2}$
are the two circular-polarization basis vectors. Apart from an overall phase factor, the transmitted angular spectrum can be written as
\begin{equation}
\widetilde{\mathbf{E}}'_p(\Theta_x,\Theta_y)
\propto
\widetilde{u}(\Theta_x,\Theta_y)
e^{-ik_x\Delta_x^{\hat{\mathbf{x}}}}
\left(
\hat{\mathbf{x}}
+k_y\Delta_y^{\hat{\mathbf{x}}}\hat{\mathbf{y}}
\right),
\label{eq:EP_x_prime}
\end{equation}
where the complex longitudinal and transverse shifts are
\begin{equation}
\Delta_x^{\hat{\mathbf{x}}}
=
-\frac{1}{k}
\frac{\partial\phi_x}{\partial\theta},
\qquad
\Delta_y^{\hat{\mathbf{x}}}
=
\frac{1}{k}
\left(1-e^{i\Phi}\right)\cot\theta.
\notag
\end{equation}
The detailed derivation is provided in Appendix~\ref{AP:SHEL}.

Under the condition
$|\Delta_x^{\hat{\mathbf{x}}}|,\ |\Delta_y^{\hat{\mathbf{x}}}|\ll w_0$,
the inverse Fourier transform gives the transmitted field in real space as
\begin{equation}
\begin{aligned}
\mathbf{E}'_p(x,y)
\propto{}&
\hat{\mathbf{e}}_{+}\,
u\left(x-\Delta_x^{\hat{\mathbf{x}}},y-\Delta_y^{\hat{\mathbf{x}}}\right)
+
\hat{\mathbf{e}}_{-}\,
u\left(x-\Delta_x^{\hat{\mathbf{x}}},y+\Delta_y^{\hat{\mathbf{x}}}\right).
\end{aligned}
\label{eq:realspace_x}
\end{equation}
The real part of $\Delta_x^{\hat{\mathbf{x}}}$ describes the in-plane spatial Goos--H\"anchen shift~\cite{Ziauddin2010,Ziauddin2011}, while its imaginary part corresponds to the associated angular deflection. More importantly, Eq.~(\ref{eq:realspace_x}) shows that the two circularly polarized components undergo opposite transverse shifts, $\pm\Delta_y^{\hat{\mathbf{x}}}$, along the $y$ direction. This spin-dependent splitting constitutes the SHEL.

In the following, we focus on the transverse spatial displacement and angular deflection associated with the SHEL. Decomposing the complex transverse shift as
$\Delta_y^{\hat{\mathbf{\mathbf{x}}}}
=
\Delta_y^{\hat{\mathbf{\mathbf{x}}}\prime}
+i\Delta_y^{\hat{\mathbf{x}}\prime\prime}$,
we obtain
\begin{subequations}
\label{eq:x_shifts}
\begin{align}
&\Delta_y^{\hat{\mathbf{x}}\prime}
=
\frac{1}{k}
\left(
1-e^{-\Phi''}\cos(\Phi')
\right)
\cot\theta,
\\
&\Delta_y^{\hat{\mathbf{x}}\prime\prime}
=
-\frac{1}{k}
e^{-\Phi''}\sin(\Phi')
\cot\theta.
\end{align}
\end{subequations}
Here, $\Delta_y^{\hat{\mathbf{x}}\prime}$ is the spatial transverse shift at the exit plane of the medium. $\Delta_y^{\hat{\mathbf{x}}\prime\prime}$ characterizes the angular spin Hall shift of the transmitted beam.

Similarly, for a $y$-polarized incident probe beam, the transmitted field in real space can be expressed as
\begin{equation}
\begin{aligned}
\mathbf{E}'_p(x,y)
\propto{}&
\hat{\mathbf{e}}_{+}\,
u\left(x,y-\Delta_y^{\hat{\mathbf{y}}}\right)
-
\hat{\mathbf{e}}_{-}\,
u\left(x,y+\Delta_y^{\hat{\mathbf{y}}}\right),
\end{aligned}
\label{eq:realspace_y}
\end{equation}
where
\begin{equation}
\Delta_y^{\hat{\mathbf{y}}}
=
\frac{1}{k}
\left(1-e^{-i\Phi}\right)\cot\theta
\notag
\end{equation}
is the corresponding complex transverse shift. Unlike the $x$-polarized component, the $y$-polarized component does not experience a longitudinal shift because $\phi_y$ is independent of $\theta$.
Writing
$\Delta_y^{\hat{\mathbf{y}}}
=
\Delta_y^{\hat{\mathbf{y}}\prime}
+i\Delta_y^{\hat{\mathbf{y}}\prime\prime}$,
its real and imaginary parts are
\begin{subequations}
\label{eq:y_shifts}
\begin{align}
\Delta_y^{\hat{\mathbf{y}}\prime}
&=
\frac{1}{k}
\left(
1-e^{\Phi''}\cos(\Phi')
\right)
\cot\theta,
\\
\Delta_y^{\hat{\mathbf{y}}\prime\prime}
&=
\frac{1}{k}
e^{\Phi''}\sin(\Phi')
\cot\theta.
\end{align}
\end{subequations}
Here $\Delta_y^{\hat{\mathbf{y}}\prime}$ represents the spatial transverse shift. $\Delta_y^{\hat{\mathbf{y}}\prime\prime}$ characterizes the angular spin Hall shift of the transmitted beam.
The detailed derivation is provided in Appendix~\ref{AP:SHEL}.
The expressions (\ref{eq:x_shifts}) and (\ref{eq:y_shifts}) show explicitly that the dispersive anisotropy $\Phi'$ primarily governs the sign of the angular shifts through $\sin\Phi'$, whereas the absorptive anisotropy $\Phi''$ can exponentially enhance or suppress the SHEL through the factors $e^{\pm\Phi''}$.

\section{Optical susceptibility}\label{sec:atomic susceptibilities}

The above discussion demonstrates that the tunable optical susceptibility of the atomic medium provides flexible control over the SHEL. In the following, we focus on the role of the nonlinear optical response of the probe field. For the numerical calculations, the relevant atomic states are chosen as
$|1\rangle=|5\,^2S_{1/2}, F=1, m_F =-1\rangle $, $|2\rangle= |5\,^2S_{1/2}, F=1, m_F=1\rangle $, $|0\rangle= |5\,^2S_{1/2}, F=1, m_F=0\rangle $, $|3\rangle= |5\,^2P_{3/2}, F=0\rangle$, the Rydberg state $|4\rangle= |60\,S_{1/2}\rangle$. 
The atoms are assumed to be initially prepared in the two outermost ground-state Zeeman sublevels $|1\rangle$ and $|2\rangle$, with equal populations, i.e.
$\rho_{11}^{(0)}=\rho_{22}^{(0)}={1}/{2}$.

Under the electric-dipole and rotating-wave approximations, the Hamiltonian of the interacting atomic ensemble can be written as
$\hat{H}
=
\mathcal{N}_a
\int d^3{r}\,
\hat{\mathcal{H}}(\mathbf{r},t)$ 
($d^3{r}=dx\,dy\,dz$),
where $\mathcal{N}_a$ is the atomic number density and $\hat{\mathcal{H}}(\mathbf{r},t)$ is the Hamiltonian density with the form
\begin{align}
\hat{\mathcal{H}} = & -\sum_{\alpha=2}^{4} \hbar \Delta_\alpha \hat{S}_{\alpha\alpha}(\mathbf{r},t) - \hbar\left[ \Omega_{p1}^* \hat{S}_{31}(\mathbf{r},t) + \Omega_{p2}^* \hat{S}_{32}(\mathbf{r},t) 
\right. \notag \\
 & \left. 
+\Omega_c^* \hat{S}_{43}(\mathbf{r},t) + \mathrm{H.c.} \right]\notag\\
& + \mathcal{N}_a \int d^3r' \, \hat{S}_{44}(\mathbf{r}',t) \hbar V(\mathbf{r}' - \mathbf{r}) \hat{S}_{44}(\mathbf{r},t),
\label{eq:hamiltonian}
\end{align}
where $\hat{S}_{\alpha\beta} = |\beta\rangle\langle\alpha| \exp\{i[(\mathbf{k}_\beta - \mathbf{k}_\alpha)\cdot\mathbf{r} - (\omega_\beta - \omega_\alpha + \Delta_\beta - \Delta_\alpha)t]\}$ (\(\alpha,\,\beta=1,\,2,\,3,\,4\)) is the transition operator.  
$\Omega_{p1}$, $\Omega_{p2}$, and $\Omega_{c}$  are the half Rabi frequencies of the two circularly polarized probe-field components and the control field, respectively.
The detunings $\Delta_2 = - (E_2 - E_1)/\hbar$, $\Delta_3 = \omega_p - (E_3 - E_1)/\hbar$, and $\Delta_4 = \omega_c + \omega_p - (E_4 - E_1)/\hbar$, with $E_{\alpha}$ the eigenenergy of the state $|\alpha\rangle$. 
 The last term in Eq.~(\ref{eq:hamiltonian}) describes the interaction between atoms simultaneously excited to the Rydberg state $|4\rangle$. This Rydberg-Rydberg interaction is modeled by the van der Waals potential
$V_{\mathrm{vdW}}(\mathbf{r}'-\mathbf{r})
\equiv\hbar V(\mathbf{r}'-\mathbf{r})
=-{\hbar C_6}/{|\mathbf{r}'-\mathbf{r}|^6}$,
where $C_6$ is the van der Waals dispersion coefficient.
The Rydberg-Rydberg interaction results in energy shifts and hence induces a phenomenon called Rydberg blockade~\cite{Saffman2010}, by which only one atom can be excited to Rydberg states in any spatial region of the atomic ensemble (i.e., Rydberg blockade sphere) with radius $R_b$ as shown in Fig.~\ref{fig:model}(a) [red dashed circles].

The dynamics of the atomic ensemble is governed by the optical Bloch equation
\begin{equation}
\frac{\partial\hat{\rho}}{\partial t}
=
-\frac{i}{\hbar}
\left[
\hat{H},\hat{\rho}
\right]
-
\mathcal{L}[\hat{\rho}],
\label{Bloch}
\end{equation}
where $\hat{\rho}$ is the atomic density operator and $\mathcal{L}$ is the relaxation superoperator describing spontaneous emission, incoherent population transfer, and dephasing. The one-body density-matrix elements are defined as
$\rho_{\alpha\beta}
\equiv
\left\langle
\hat{S}_{\beta\alpha}
\right\rangle$,~
($\alpha,\,\beta=1-4$).
The explicit form of Eq.~(\ref{Bloch}) is provided in Eq.~(\ref{eqn1}) of Appendix~\ref{appendixA}.

Because the polarization-dependent refractive indices are determined by the corresponding optical susceptibilities, we next derive the optical response of the two probe-field components. The susceptibilities are directly related to the optical coherences $\rho_{31}$ and $\rho_{32}$ through~\cite{Bai2016OE,Mu2021PRA,Hang2018Parity}
\begin{subequations}
\label{eq:susceptibility}
\begin{align}
\chi_{1}=\frac{\mathcal{N}_a\left({\mathbf e}_{-}
\cdot\mathbf{p}_{13}\right)\rho_{31}}{
\varepsilon_0\mathcal{E}_{p1}},\\
\chi_{2}=\frac{\mathcal{N}_a\left({\mathbf e}_{+}
\cdot\mathbf{p}_{23}\right)\rho_{32}}
{\varepsilon_0\mathcal{E}_{p2}}.
\end{align}
\end{subequations}
Here, $\chi_1$ and $\chi_2$ denote the optical susceptibilities associated with the $\sigma^+$ and $\sigma^-$ components of the probe field, whose respective amplitudes are $\mathcal{E}_{p1}$ and $\mathcal{E}_{p2}$. The quantities $\mathbf{p}_{13}$ and $\mathbf{p}_{23}$ are the electric-dipole matrix elements associated with the transitions $|1\rangle\leftrightarrow|3\rangle$ and $|2\rangle\leftrightarrow|3\rangle$, respectively. The total optical susceptibility experienced by the probe field is given by $\chi=\chi_1+\chi_2$. 

We are interested in the steady-state response of the atomic medium. Accordingly, the time derivatives in Eq.~(\ref{Bloch}) can be neglected when the probe pulses are sufficiently long compared with the characteristic atomic relaxation times. Furthermore, when the probe fields are weak compared with the control field, Eq.~(\ref{Bloch}) can be solved perturbatively using the method of multiple scales~\cite{Newell1980}. The detailed derivation is presented in Appendix~\ref{appendixB}.

The susceptibility of the $j$th probe-polarization component, with $j=1,2$, can then be written as
\begin{align}
\chi_j=&
\chi_j^{(\mathrm{lin})}
+
\chi_{j1}^{(3,\mathrm{loc})}
|\mathcal{E}_{p1}|^2
+
\chi_{j2}^{(3,\mathrm{loc})}
|\mathcal{E}_{p2}|^2\notag
\\
&+
\chi_{j1}^{(3,\mathrm{nloc})}
|\mathcal{E}_{p1}|^2
+
\chi_{j2}^{(3,\mathrm{nloc})}
|\mathcal{E}_{p2}|^2,\label{eq:chi_total}
\end{align}
where $\chi_j^{(\mathrm{lin})}$ is the linear susceptibility, whereas $\chi_{jl}^{(3,\mathrm{loc})}$ and $\chi_{jl}^{(3,\mathrm{nloc})}$ denote the local and nonlocal third-order nonlinear susceptibilities, respectively. The cases $j=l$ and $j\neq l$ correspond to the self-Kerr and cross-Kerr responses, describing the self-phase and cross-phase modulation of the probe-field components.
The explicit expressions for
$\chi_j^{(\mathrm{lin})}$,
$\chi_{jl}^{(3,\mathrm{loc})}$, and
$\chi_{jl}^{(3,\mathrm{nloc})}$
are provided in Appendix~\ref{AP:CHI}. An important feature of the nonlinear response is its distinct dependence on the atomic density:
\begin{equation}
\chi_{jl}^{(3,\mathrm{loc})}
\propto
\mathcal{N}_a,
\qquad
\chi_{jl}^{(3,\mathrm{nloc})}
\propto
\mathcal{N}_a^2.\notag
\end{equation}
The quadratic density scaling of the nonlocal susceptibility stems from Rydberg-mediated two-atom correlations, enabling the nonlocal Kerr response to 
significantly exceed its local counterpart as the atomic density increases.

\begin{figure}[t]
\centering
\includegraphics[width=\columnwidth]{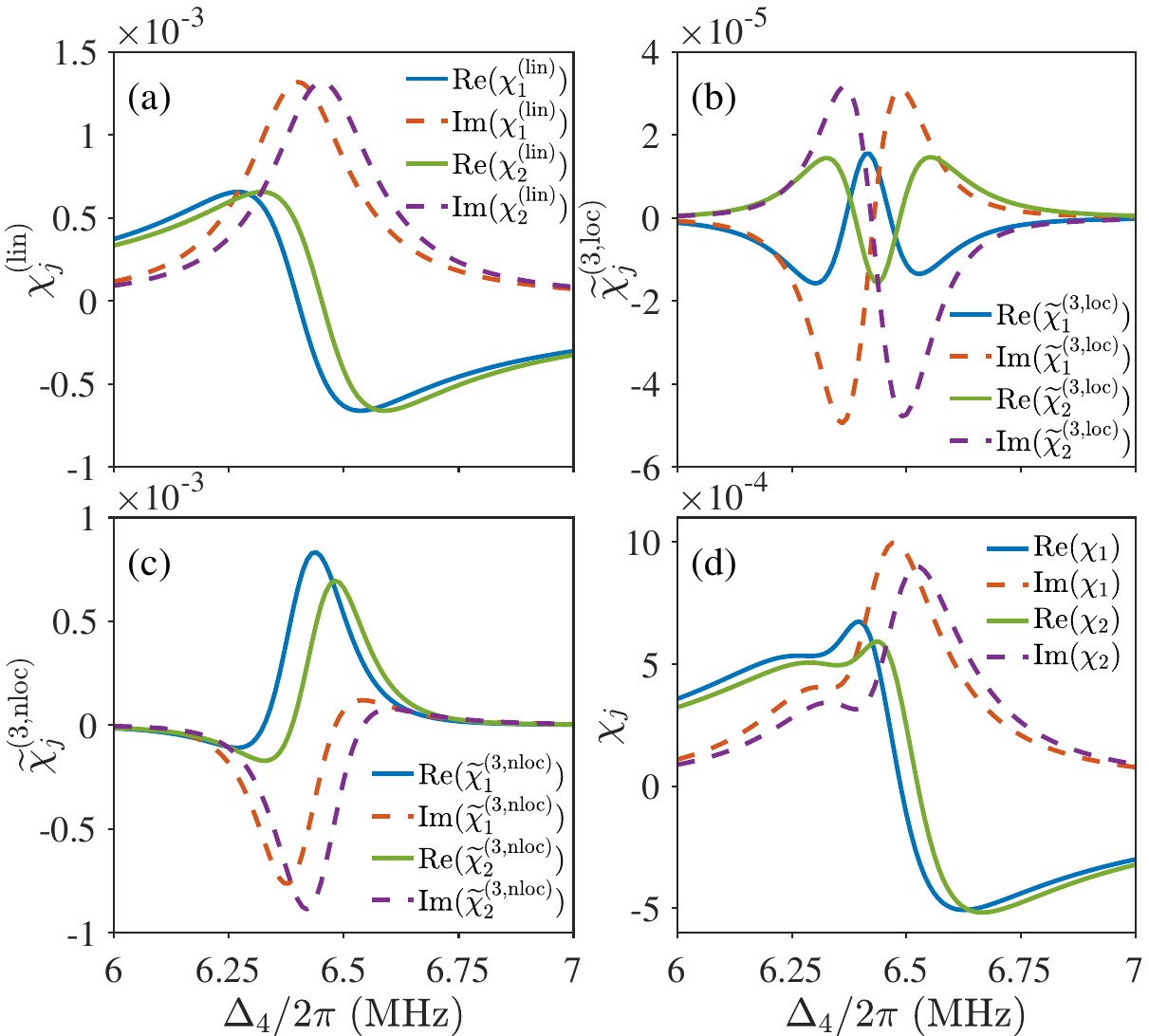}\caption{
Optical susceptibilities of the probe field as functions of the two-photon detuning $\Delta_4$. (a) Linear susceptibilities $\chi_{1}^{(\mathrm{lin})}$ and $\chi_{2}^{(\mathrm{lin})}$. (b) Local third-order nonlinear contributions $\widetilde{\chi}_{1}^{(3,\mathrm{loc})}$ and $\widetilde{\chi}_{2}^{(3,\mathrm{loc})}$. Each contribution includes both the self-Kerr and cross-Kerr responses and is defined as
$\widetilde{\chi}_{j}^{(3,\mathrm{loc})}
=\chi_{j1}^{(3,\mathrm{loc})}|\mathcal{E}_{p1}|^2
+\chi_{j2}^{(3,\mathrm{loc})}|\mathcal{E}_{p2}|^2$.
(c) Nonlocal third-order nonlinear contributions
$\widetilde{\chi}_{1}^{(3,\mathrm{nloc})}$ and
$\widetilde{\chi}_{2}^{(3,\mathrm{nloc})}$, defined as
$\widetilde{\chi}_{j}^{(3,\mathrm{nloc})}
=\chi_{j1}^{(3,\mathrm{nloc})}|\mathcal{E}_{p1}|^2
+\chi_{j2}^{(3,\mathrm{nloc})}|\mathcal{E}_{p2}|^2$.
(d) Total susceptibilities $\chi_1$ and $\chi_2$, obtained by combining the linear, local nonlinear, and nonlocal nonlinear contributions. The indices $j=1$ and $2$ correspond to the $\sigma^+$ and $\sigma^-$ components of the probe field, respectively. The blue solid and red dashed curves represent the real and imaginary parts of the susceptibility for the $\sigma^+$ component, respectively, whereas the green solid and purple dashed curves represent the corresponding real and imaginary parts for the $\sigma^-$ component.
}
\label{fig:susceptibility}
\end{figure}

To elucidate the enhancement and tunable control of the SHEL in the Rydberg atomic medium, we compare the linear, local nonlinear, and nonlocal nonlinear contributions to the optical susceptibilities. Before presenting the relation between the optical susceptibilities and  the two-photon detuning $\Delta_4$. The relevant relaxation parameters are set to $\Gamma_{34}=2\pi\times16.7~\mathrm{kHz}$,
$\Gamma_{13}=\Gamma_{23}=
{\Gamma_3}/{2}=2\pi\times3.03~\mathrm{MHz}$,
and $\Gamma_{12}=\Gamma_{21}=2\pi\times1.0~\mathrm{kHz}$.
For the Rydberg state with principal quantum number $n=60$, the van der Waals coefficient is chosen as $C_6\simeq-2\pi\times140~\mathrm{GHz}\,\mu\mathrm{m}^{6}$~\cite{Singer2005JPhysB}. Other parameters
$\Omega_c=2\pi\times32~\mathrm{MHz}$,
$\Delta_2=2\pi\times0.05~\mathrm{MHz}$,
$\Delta_3=2\pi\times160~\mathrm{MHz}$,
$\lambda_p=780~\mathrm{nm}$,
$\mathbf{p}_{13}=\mathbf{p}_{23}
=3.58\times10^{-29}~\mathrm{C\cdot m}$,
$\mathcal{E}_{p1}=\mathcal{E}_{p2}=0.56~\mathrm{V/m}$,
$\mathcal{N}_a=4\times10^{10}~\mathrm{cm}^{-3}$.

As shown in Fig.~\ref{fig:susceptibility}(a), the linear susceptibility $\chi_j^{(\mathrm{lin})}$ exhibits a characteristic resonant response. For the $\sigma^+$ component, the imaginary part $\operatorname{Im}[\chi_1^{(\mathrm{lin})}]$,  represented by the red dashed line, displays a pronounced positive peak near $\Delta_4=2\pi\times6.4~\mathrm{MHz}$. This peak approximately satisfies the resonance condition
$\Delta_4\simeq {|\Omega_c|^2}/{\Delta_3}$
and signifies strong resonant absorption~\cite{WuJH2022PRA}. Meanwhile, the real part $\operatorname{Re}[\chi_1^{(\mathrm{lin})}]$, shown by the blue solid line, varies rapidly and changes sign across the resonance, revealing a steep anomalous-dispersion response. The corresponding real and imaginary parts of $\chi_2^{(\mathrm{lin})}$ exhibit similar spectral profiles but do not completely overlap with those of $\chi_1^{(\mathrm{lin})}$ because of the finite ground-state detuning $\Delta_2$. In the limit $\Delta_2=0$, the two polarization components become nearly degenerate, and their susceptibility curves almost coincide. The real and imaginary parts of $\chi_2$ are presented by green solid and purple dashed lines.

Figure~\ref{fig:susceptibility}(b) shows the local third-order nonlinear contributions to the optical susceptibilities. Although the difference between the two polarization components becomes more pronounced than in the linear response, the local nonlinear contributions $\widetilde{\chi}_{j}^{(3,\mathrm{loc})}=\Sigma_{l=1}^2\chi_{jl}^{(3,\mathrm{loc})}|\mathcal{E}_{pl}|^2$ remain nearly two orders of magnitude smaller than the corresponding linear susceptibilities. The local Kerr response therefore produces only a minor modification of the overall optical properties under the parameters considered.

A markedly stronger nonlinear response is observed in Fig.~\ref{fig:susceptibility}(c). The nonlocal third-order contributions originate from long-range Rydberg--Rydberg interactions and substantially exceed their local counterparts. Consequently, the terms $\widetilde{\chi}_{j}^{(3,\mathrm{nloc})}=\Sigma_{l=1}^2\chi_{jl}^{(3,\mathrm{nloc})}|\mathcal{E}_{pl}|^2$ dominate the nonlinear modification of both the dispersive and absorptive responses. Figure~\ref{fig:susceptibility}(d) presents the total susceptibilities obtained by combining the linear, local nonlinear, and nonlocal nonlinear contributions. The linear response establishes the underlying dispersive and absorptive background, whereas the Rydberg-mediated nonlocal nonlinearity strongly reshapes the overall spectral response. This comparison identifies the nonlocal third-order susceptibility as the principal nonlinear mechanism for enhancing and controlling the SHEL in the present system.

\section{Giant and sign-tunable spin Hall effect of light}\label{sec:IV}

\begin{figure}
\centering
\includegraphics[width=\columnwidth]{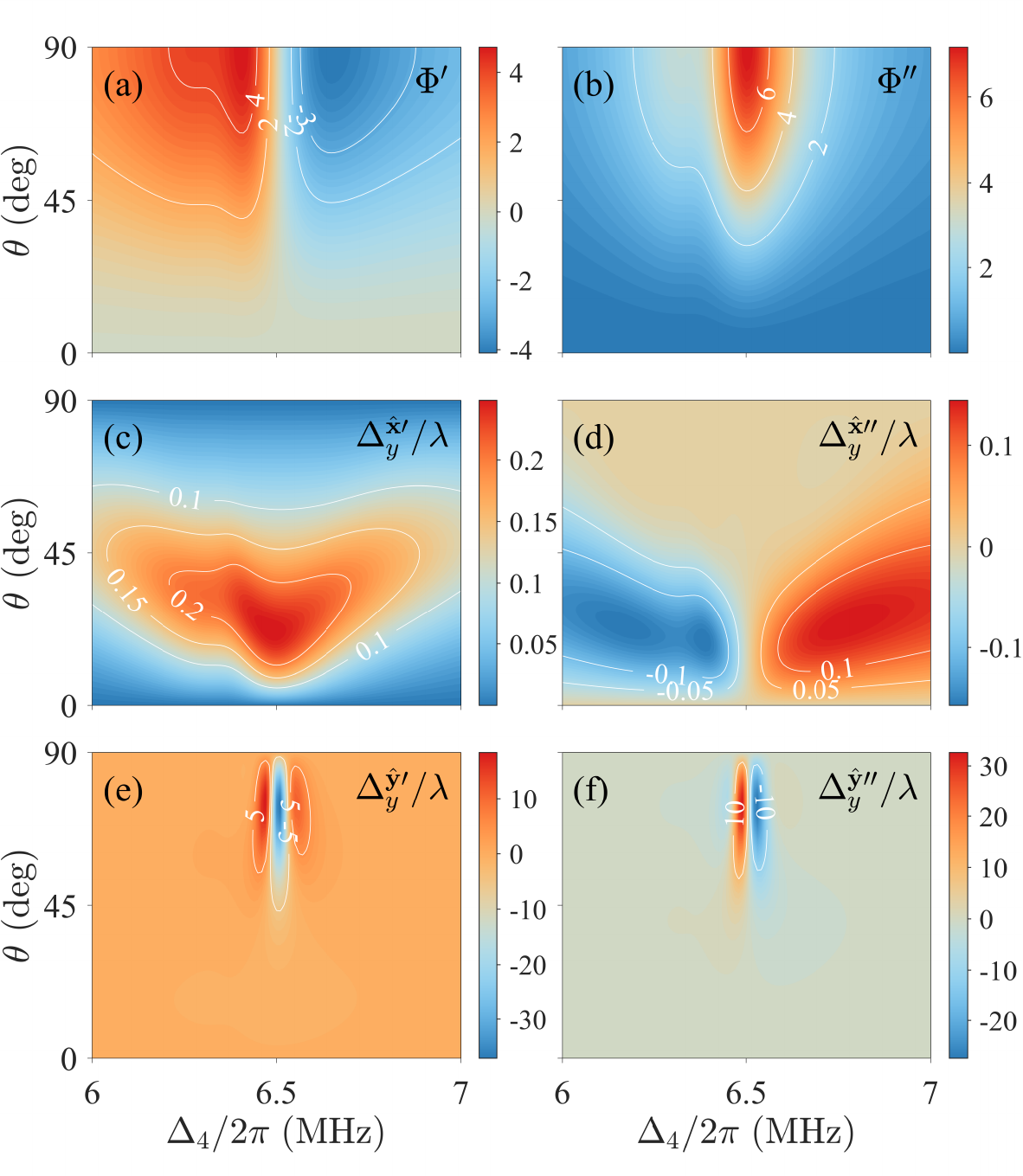}
\caption{
Complex phase difference and spin-dependent transverse shifts as functions of the two-photon detuning $\Delta_4$ and the orientation angle $\theta$. Panels (a) and (b) show the real and imaginary parts, $\Phi'$ and $\Phi''$, of the complex phase difference $\Phi$, respectively. Panels (c) and (d) show the normalized spatial shift $\Delta_y^{\hat{\mathbf{x}}\prime}/\lambda$ and angular shift parameter $\Delta_y^{\hat{\mathbf{x}}\prime\prime}/\lambda$, respectively, for an $x$-polarized incident probe beam. Panels (e) and (f) show the corresponding quantities, $\Delta_y^{\hat{\mathbf{y}}\prime}/\lambda$ and $\Delta_y^{\hat{\mathbf{y}}\prime\prime}/\lambda$, for a $y$-polarized incident probe beam. Here, $\lambda=\lambda_p$, and all other parameters are the same as those used in Fig.~\ref{fig:susceptibility}.
}
\label{fig:shel}
\end{figure}

\subsection{Polarization asymmetry of SHEL}

In this section, we investigate the spin-dependent transverse shifts,
focusing on the
 role of the Rydberg-mediated nonlocal Kerr nonlinearity in enhancing and controlling the SHEL. Equations~(\ref{eq:ny}) and~(\ref{eq:nx}) show that the two orthogonal probe-polarization components experience different refractive indices. This optically induced anisotropy, together with the spin-redirection geometric phase acquired by the angular-spectrum components of the probe beam, gives rise to the SHEL. Because the anisotropy depends strongly on the angle $\theta$ between the quantization axis and the $z$ axis, the resulting spin-dependent transverse shifts can be controlled by varying both $\theta$ and the two-photon detuning $\Delta_4$. Figure~\ref{fig:shel} presents the corresponding dependence of the phase anisotropy and transverse shifts on these two parameters.

As shown in Fig.~\ref{fig:shel}(a), the real part of the phase difference, $\Phi'$, characterizes the dispersive anisotropy of the atomic medium. It varies rapidly near resonance and changes sign as $\Delta_4$ is tuned through approximately $2\pi\times6.5~\mathrm{MHz}$. On the low-detuning side of this point, the magnitude of the dispersive anisotropy increases with $\theta$, whereas on the high-detuning side, it decreases as $\theta$ increases. The imaginary part $\Phi''$, shown in Fig.~\ref{fig:shel}(b), quantifies the absorptive anisotropy. It remains positive throughout the parameter region considered and reaches a pronounced maximum near $\Delta_4\simeq2\pi\times6.5~\mathrm{MHz}$, where the differential absorption experienced by the two orthogonal polarization components is strongly enhanced. Moreover, $\Phi''$ increases monotonically with $\theta$, indicating that tilting the quantization axis farther away from the probe field propagation direction strengthens the absorptive anisotropy.

Figures~\ref{fig:shel}(c) and~\ref{fig:shel}(d) show the normalized spatial shift $\Delta_y^{\hat{\mathbf{x}}\prime}/\lambda$ and angular-shift parameter $\Delta_y^{\hat{\mathbf{x}}\prime\prime}/\lambda$, respectively, for an $x$-polarized incident probe beam. The spatial shift is predominantly positive and reaches its maximum near resonance at an intermediate value of $\theta$. By contrast, $\Delta_y^{\hat{\mathbf{x}}\prime\prime}/\lambda$ changes sign across the resonant region. According to Eq.~(\ref{eq:x_shifts}), this reversal is governed primarily by the sign change of $\sin\Phi'$. The actual angular deflection of the transmitted beam is given by $\Delta_y^{\hat{\mathbf{x}}\prime\prime}/z_R$, where $z_R$ is the Rayleigh range.

A substantially stronger response is obtained for a $y$-polarized incident probe beam. As shown in Fig.~\ref{fig:shel}(e), the normalized spatial shift $\Delta_y^{\hat{\mathbf{y}}\prime}/\lambda$ exhibits a much sharper and more pronounced resonant feature than its $x$-polarized counterpart. This enhancement originates primarily from the factor $e^{\Phi''}$ in Eq.~(\ref{eq:y_shifts}), which exponentially amplifies the influence of the absorptive anisotropy on the transverse displacement. The strongest spin-dependent beam splitting therefore occurs in the resonant region, where the dispersive response varies rapidly and the absorptive anisotropy is strongly enhanced.
A similar enhancement is observed for the angular-shift parameter $\Delta_y^{\hat{\mathbf{\mathbf{y}}}\prime\prime}/\lambda$ in Fig.~\ref{fig:shel}(f). Its magnitude increases markedly near resonance because of the same factor $e^{\Phi''}$. In addition, $\Delta_y^{\hat{\mathbf{y}}\prime\prime}$ reverses sign in the vicinity of $\Delta_4=2\pi\times6.5~\mathrm{MHz}$, with the reversal again determined predominantly by the sign change of $\sin\Phi'$. The corresponding physical angular deflection is $\Delta_y^{\hat{\mathbf{y}}\prime\prime}/z_R$.

Overall, Fig.~\ref{fig:shel} reveals a pronounced polarization dependence of the SHEL. Both the spatial displacement and angular deflection are generally much larger for the $y$-polarized incident beam than for the $x$-polarized beam. 
This asymmetry stems from the opposite exponential factors governing the two polarizations: the $x$-polarized shifts contain the suppressing factor $e^{-\Phi''}$, whereas the $y$-polarized shifts contain the enhancing factor $e^{\Phi''}$.
This polarization-dependent enhancement can be understood via the Jones matrix,
where the effect relates to the transmission and the phase-dependent geometric contribution.
These results demonstrate that both the magnitude and direction of the SHEL can be efficiently controlled through the joint tuning of the two-photon detuning $\Delta_4$ and the orientation angle $\theta$. 


\begin{figure}
\centering
\includegraphics[width=\columnwidth]{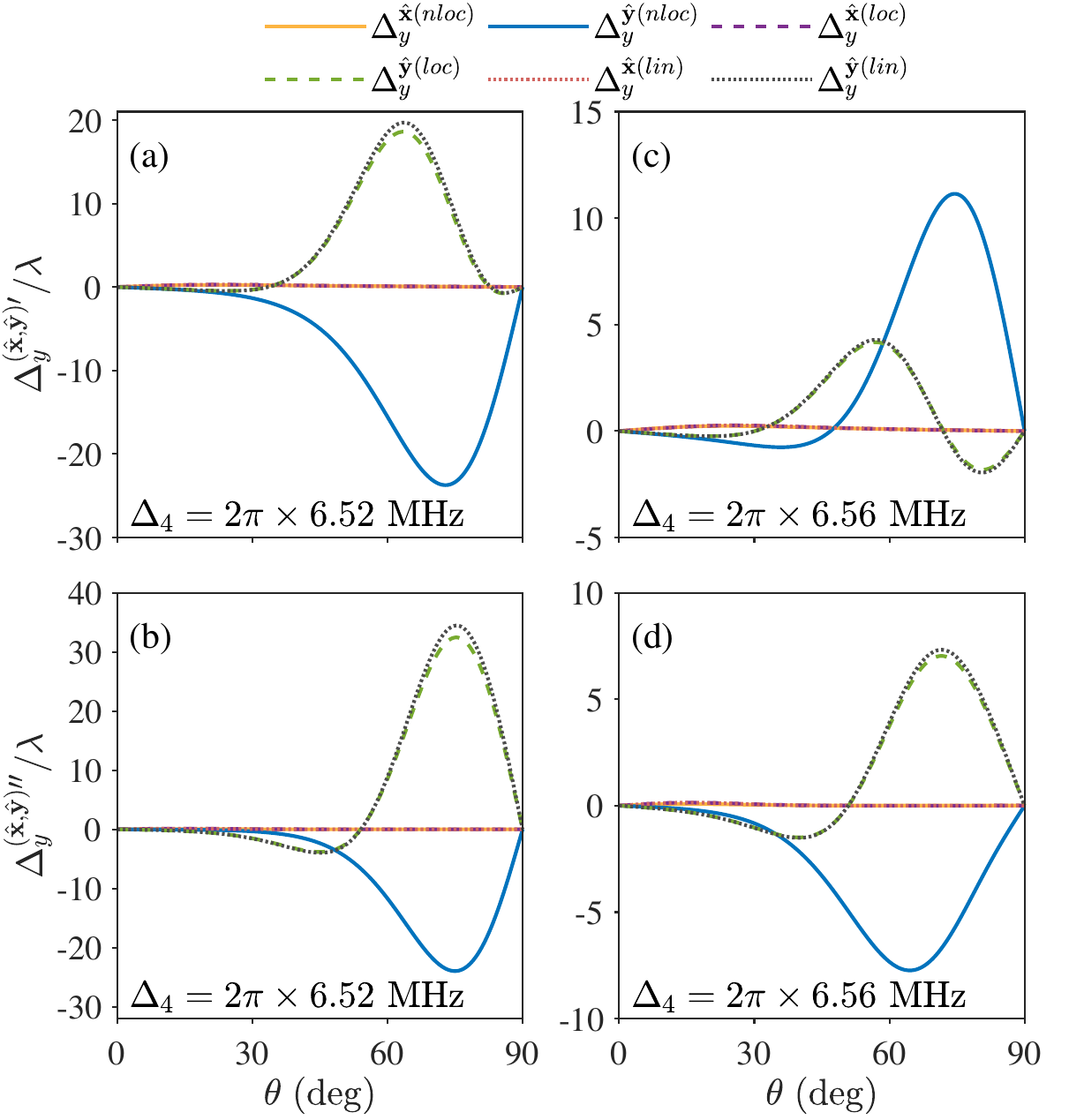}
\caption{Spatial and angular shifts associated with the SHEL as functions of the angle $\theta$ between the quantization axis and the $z$ axis. Panels (a) and (c) show the normalized spatial shifts $\Delta_y^{(\hat{\mathbf{x}},\hat{\mathbf{y}})\prime}/\lambda$, whereas panels (b) and (d) show the normalized angular shift parameters $\Delta_y^{(\hat{\mathbf{x}},\hat{\mathbf{y}})\prime\prime}/\lambda$. For the $y$-polarized incident beam, the black dotted, green dashed, and blue solid lines correspond to the linear response alone, the combined linear and local nonlinear responses, and the total response including the nonlocal nonlinearity, respectively. For the $x$-polarized incident beam, the corresponding three cases are represented by the red dotted, purple dashed, and yellow solid lines, respectively. The detuning is fixed at $\Delta_4=2\pi\times6.52~\mathrm{MHz}$ in panels (a) and (b), and at $\Delta_4=2\pi\times6.56~\mathrm{MHz}$ in panels (c) and (d). All other parameters are the same as those used in Fig.~\ref{fig:susceptibility}.
}
\label{fig:SHEL_shift}
\end{figure}

\subsection{Detuning and angle controlled SHEL}

The above discussion demonstrates that the SHEL in the Rydberg atomic medium is highly sensitive to the two-photon detuning $\Delta_4$. Pronounced transverse shifts occur predominantly in the strongly absorptive  region and are substantially larger for the $y$-polarized component than for the $x$-polarized component. To further elucidate the role of the nonlocal nonlinear response near $\Delta_4=2\pi\times6.5~\mathrm{MHz}$, Fig.~\ref{fig:SHEL_shift} compares the spatial and angular shifts obtained by successively including the linear, local nonlinear, and nonlocal nonlinear susceptibilities as functions of the angle $\theta$ between the quantization axis and the $z$ axis. Figures~\ref{fig:SHEL_shift}(a) and~(b) present the spatial and angular shifts, respectively, at the fixed detuning $\Delta_4=2\pi\times6.52~\mathrm{MHz}$.

For the $x$-polarized component, both the spatial and angular shifts remain negligibly small over the entire angular range, regardless of whether the nonlocal nonlinear contribution is included, consistent with Figs.~\ref{fig:shel}(c) and (d). In sharp contrast, the $y$-polarized component exhibits a pronounced dependence on the different susceptibility contributions. As shown in Fig.~\ref{fig:SHEL_shift}(a), the spatial shift remains positive over most of the angular range and reaches a maximum near $\theta\approx65^\circ$ when only the linear response is considered, see black dotted line. Including the local nonlinear susceptibility produces almost no discernible change (green dashed line), confirming that the local nonlinear contribution has a negligible influence on both the spatial and angular shifts. Once the nonlocal nonlinearity is included (blue solid line), however, the spatial shift is dramatically enhanced in the negative direction, reaching  a pronounced extreme near $\theta\approx75^\circ$. This reversal indicates that the nonlocal Rydberg--Rydberg interaction controls not only the magnitude of the SHEL but also the direction of the spin-dependent transverse displacement. A similar trend is observed for the angular shift in Fig.~\ref{fig:SHEL_shift}(b): it is positive in the absence of the nonlocal contribution but becomes negative after the nonlocal nonlinear response is taken into account. These results identify the nonlocal Kerr nonlinearity as the dominant mechanism underlying the enhancement and directional reversal of the SHEL in the resonant regime.

For $\Delta_4=2\pi\times6.56~\mathrm{MHz}$, the corresponding spatial and angular shifts are shown in Figs.~\ref{fig:SHEL_shift}(c) and~\ref{fig:SHEL_shift}(d), respectively. As in Figs.~\ref{fig:SHEL_shift}(a) and~\ref{fig:SHEL_shift}(b), both shifts remain negligibly small for the $x$-polarized incident beam. By contrast, the $y$-polarized component exhibits a pronounced SHEL response, with both shifts displaying strong and distinct dependence on $\theta$.

As shown in Fig.~\ref{fig:SHEL_shift}(c), when the nonlocal nonlinear response is included, the spatial shift of the $y$-polarized component remains small at low values of $\theta$, but then increases rapidly and reaches a pronounced positive maximum near $\theta\approx75^\circ$. This behavior contrasts sharply with that at $\Delta_4=2\pi\times6.52~\mathrm{MHz}$ [Fig.~\ref{fig:SHEL_shift}(a)], where the nonlocal nonlinearity enhanced spatial shift is negative. Therefore, even a slight variation in the detuning can reverse the direction of the spatial displacement induced by the nonlocal nonlinearity. In comparison, when the nonlocal contribution is excluded, the spatial shift is substantially smaller and varies only moderately with $\theta$.

Figure~\ref{fig:SHEL_shift}(d) further shows that the angular shift of the $y$-polarized component is strongly reshaped by the nonlocal nonlinear response. Upon including the nonlocal contribution, the angular shift becomes negative and attains a large absolute value over the range $\theta\approx70^\circ$--$80^\circ$. Without the nonlocal contribution, however, it remains positive within the same angular interval. These results demonstrate that the nonlocal third-order susceptibility is the dominant mechanism governing both the enhancement and the directional reversal of the angular deflection.

\subsection{Nonlocal nonlinearity enhanced SHEL}

\begin{figure}
\centering
\includegraphics[width=\columnwidth]{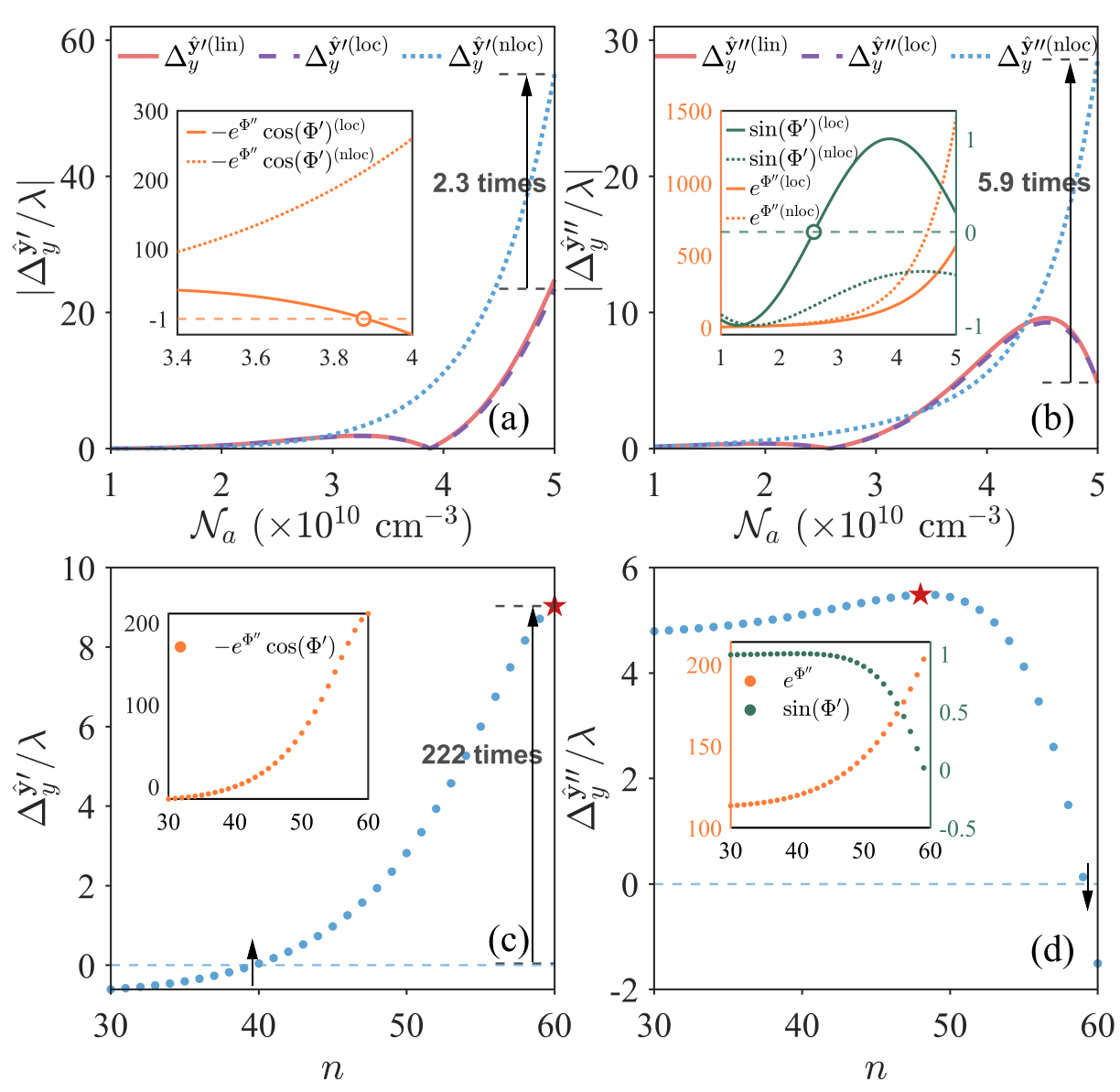}\caption{
Spatial and angular shifts of the $y$-polarized component as functions of the atomic density $\mathcal{N}_a$ and the principal quantum number $n$. Panels (a) and (b) show the density dependence of the normalized spatial shift $\Delta_y^{\hat{\mathbf{y}}\prime}/\lambda$ and angular-shift parameter $\Delta_y^{\hat{\mathbf{y}}\prime\prime}/\lambda$, respectively, whereas panels (c) and (d) show their corresponding dependence on $n$. The red solid, purple dashed, and blue dotted curves represent the linear response alone, the combined linear and local nonlinear responses, and the total response including the nonlocal nonlinear susceptibility, respectively. The insets of panels (a) and (c) show the phase-dependent spatial-shift factor $-e^{\Phi''}\cos(\Phi')$, represented by the orange solid and dotted curves. The insets of panels (b) and (d) show the phase factor $\sin(\Phi')$ and the absorptive enhancement factor $e^{\Phi''}$, represented by the green solid and dotted curves  and the orange solid and dotted curves, respectively. The detuning is fixed at $\Delta_4=2\pi\times6.56~\mathrm{MHz}$ in panels (a) and (b), and at $\Delta_4=2\pi\times6.57~\mathrm{MHz}$ in panels (c) and (d). Except for the detuning specified above, all other parameters are the same as those used in Figs.~\ref{fig:SHEL_shift}(c) and~(d).
}
\label{fig:SHEL_shift_3}
\end{figure}

Because the nonlocal nonlinear susceptibility originates from long-range Rydberg--Rydberg interactions, its strength can be systematically controlled through experimentally accessible parameters, such as the atomic density $\mathcal{N}_a$ and the principal quantum number $n$. This tunability provides an effective means of manipulating the SHEL in the Rydberg atomic medium. Figure~\ref{fig:SHEL_shift_3} shows the magnitudes of the spatial and angular shifts as functions of $\mathcal{N}_a$ and the principal quantum number $n$ of the Rydberg state.

As shown in Fig.~\ref{fig:SHEL_shift_3}(a), the red solid and purple dashed curves exhibit nearly identical density dependence over the range considered, confirming that the local Kerr response has only a negligible influence on the spatial shift. The inset shows the corresponding variation of the phase-dependent term
$-e^{\Phi^{\prime\prime}}\cos(\Phi')$
for the responses with and without the nonlocal contribution. According to Eq.~(\ref{eq:y_shifts}), the spatial shift vanishes when
$e^{\Phi''}\cos(\Phi')=1$.
Thus, the zero crossing marked by the orange circle in the inset arises from the cancellation between the constant term and the phase dependent contribution. Across this point, the sign of $1-e^{\Phi''}\cos(\Phi')$ changes, leading to a reversal in the direction of the spatial displacement.

Once the nonlocal nonlinear susceptibility is included, the resulting spatial shift deviates markedly from those obtained using only the linear and local nonlinear responses. In particular, the shift increases rapidly at high atomic densities and reaches several tens of wavelengths. At
$\mathcal{N}_a=5\times10^{10}~\mathrm{cm}^{-3}$,
the spatial shift obtained in the presence of the nonlocal nonlinearity is approximately $2.3$ times larger than that obtained without the nonlocal contribution. Its magnitude reaches
\begin{equation}
\left|
\Delta_y^{\hat{\mathbf{y}}\prime(\mathrm{nloc})}
\right|
\simeq
55\lambda
\simeq
42.9~\mu\mathrm{m}.\notag
\end{equation}
This substantial enhancement highlights the dominant role of Rydberg-mediated nonlocal interactions in amplifying the spatial spin Hall displacement.

A similar density dependent enhancement is observed for the angular shift in Fig.~\ref{fig:SHEL_shift_3}(b). The close agreement between the red solid and purple dashed curves again demonstrates that the local nonlinear response contributes only weakly to the angular SHEL. The inset of Fig.~\ref{fig:SHEL_shift_3}(b) shows the phase dependent factor governing the angular response. Because
$\Delta_y^{\hat{\mathbf{y}}\prime\prime}\propto e^{\Phi''}\sin\Phi'$,
and $e^{\Phi''}$ is always positive, the zero crossings of the angular shift are determined by
$\sin(\Phi')=0$.
The zero crossing marked by the green circle therefore results from the phase condition imposed by $\Phi'$, and the associated sign change reverses the direction of the angular deflection.

When the nonlocal nonlinear susceptibility is included, the angular shift is strongly enhanced over specific density intervals, with the enhancement becoming particularly pronounced at high atomic densities. At $\mathcal{N}_a=5\times10^{10}~\mathrm{cm}^{-3}$, the angular shift is approximately $5.9$ times larger than that obtained without the nonlocal contribution. These results demonstrate that the atomic density provides an efficient control parameter for tailoring the Rydberg-mediated nonlocal optical response and, consequently, both the spatial and angular manifestations of the SHEL.

Figures~\ref{fig:SHEL_shift_3}(c) and (d) show the spatial and angular transverse shifts, respectively, as functions of the principal quantum number $n$. As shown in Fig.~\ref{fig:SHEL_shift_3}(c), the magnitude of the spatial shift increases monotonically with $n$, with the enhancement becoming particularly pronounced for highly excited Rydberg states.
We find that the spatial shift reverse sign when the principal quantum number of the Rydberg state are around $n=40$. The black arrow indicate the reversal of direction in Fig.~\ref{fig:SHEL_shift_3}(c).
 At $n=60$, the spatial shift exceeds $9\lambda$, an enhancement by a factor of $222$ (more than two orders of magnitude) over the value obtained at the lower Rydberg level $n=40$.

To elucidate the physical origin of this enhancement, the inset of Fig.~\ref{fig:SHEL_shift_3}(c) shows the phase dependent contribution $-e^{\Phi''}\cos(\Phi')$,
which, together with the constant term, determines the spatial-shift factor
$1-e^{\Phi''}\cos(\Phi')$.
As $n$ increases, the composite factor $e^{\Phi''}\cos(\Phi')$ varies substantially because both the dispersive phase difference $\Phi'$ and the absorptive phase difference $\Phi''$ are modified by the increasingly strong Rydberg-mediated nonlocal response. Consequently, the magnitude of $1-e^{\Phi''}\cos(\Phi')$ grows rapidly at large $n$, giving rise to the pronounced enhancement of the spatial displacement.

Figure~\ref{fig:SHEL_shift_3}(d) presents the angular shift as a function of $n$. In contrast to the monotonic behavior of the spatial shift, the angular shift initially increases, reaches a maximum near $n=48$, and then decreases rapidly, eventually becoming negative at sufficiently large $n$, around $n=60$. The black arrow indicate the reversal of direction. This sign reversal signifies a reversal in the direction of the spin-dependent angular deflection. The inset of Fig.~\ref{fig:SHEL_shift_3}(d) shows the corresponding variations of $e^{\Phi^{\prime\prime}}$ and $\sin(\Phi^{\prime})$. With increasing $n$, the amplitude factor $e^{\Phi^{\prime\prime}}$ increases continuously, whereas the phase factor $\sin(\Phi^{\prime})$ decreases and 
ultimately changes sign.
 The competition between these two factors gives rise to the nonmonotonic behavior and subsequent reversal of the angular shift. These results demonstrate that the principal quantum number provides an effective control parameter for tailoring both the magnitude and direction of the SHEL. Because the long-range Rydberg--Rydberg interaction depends strongly on $n$, access to higher-lying Rydberg states substantially strengthens the nonlocal nonlinear response and thereby enables flexible control of the spin-dependent transverse shifts.

Overall, our results establish the nonlocal Kerr nonlinearity arising from long-range Rydberg--Rydberg interactions as a key mechanism for enhancing and controlling the SHEL in a cold Rydberg atomic medium. The nonlocal response can substantially alter both the magnitude and sign of the spatial and angular shifts, particularly in the vicinity of atomic resonance. Moreover, the SHEL can be manipulated through experimentally accessible parameters, including the atomic density, optical detuning, orientation of the quantization axis, and principal quantum number of the Rydberg state. Cold Rydberg atomic ensembles therefore constitute a versatile platform for controlling spin-dependent light propagation, with potential applications in precision optical sensing and spin-controlled photonic devices.

\section{CONCLUSIONS}\label{sec:V}

In this work, we have investigated the enhancement and tunable control of the SHEL in a Rydberg atomic medium by analyzing its optical response and spin dependent transverse shifts. The SHEL exhibits a strong dependence on both the two-photon detuning $\Delta_4$ and the angle $\theta$ between the quantization axis and the $z$ axis. Tuning these two parameters jointly enables flexible control over the magnitude and direction of the polarization dependent transverse shifts. In particular, choosing a $y$-polarized rather than an $x$-polarized incident beam enhances both the spatial and angular shifts by factors of several tens and, under appropriate conditions, by up to two orders of magnitude. 
This polarization-dependent enhancement 
occurs in transmission and arises from the phase-dependent geometric contribution.

We have further compared the transverse shifts obtained with and without Rydberg--Rydberg interactions. The results demonstrate that the nonlocal third-order susceptibility arising from these long-range interactions plays a dominant role in reshaping the SHEL. Incorporating the nonlocal nonlinear response can substantially enhance the transverse shifts and even reverse their signs, thereby controlling not only the magnitude of the SHEL but also the direction of the spin-dependent displacement and angular deflection.
Because nonlocal nonlinear susceptibility is closely related to the atomic density and the Rydberg excitation level, we have also examined the dependence of the SHEL on the atomic density $\mathcal{N}_a$ and the principal quantum number $n$. The results show that suitable choices of $\mathcal{N}_a$ and $n$ can markedly enhance the spatial and angular shifts. 



Notably, in our proposed model the spatial transverse shift can reach several tens of
micrometers, so that the separation between the two spin components
approaches $0.1$~mm ($0.086$~mm for our parameters). A displacement of
$0.1$~mm between the two reflected spots, corresponding to the two
polarization states, is resolvable by the naked eye. Consequently, the shift
can be detected directly with a CCD camera~\cite{PhysRevLett.124.053902}, obviating the need for weak-measurement techniques~\cite{Hosten2008}.
These findings establish Rydberg-mediated nonlocal optical nonlinearity as an
efficient and tunable mechanism for manipulating spin-dependent light
transport in atomic media.


\begin{acknowledgments}
L. Q., Y. Z., and S. Z. acknowledge the support of National Natural Science Foundation of China (Grant No. 12404377, No. 12304357, and No. 62505079),  and the Program of China Scholarship Council under Grants No. 202508410217.
X. Z. acknowledges the support from the Key Scientific Research Project of Colleges and Universities in Henan Province (Grant No. 26B140007) and Natural Science Foundation of Henan Province (Grant No. 252300421995). Z. Z. acknowledges the support from the Key International Cooperation Project in Henan Province (Grant No. 261111521200). W. L. acknowledges financial support from UKRI UKRI3702 (QCS) and the University of Nottingham. 
\end{acknowledgments}

\appendix

\section{Explicit expression of spin Hall effect of light}
\label{AP:SHEL}

The transmitted field $\widetilde{\mathbf{E}}'$ is obtained by propagating each plane-wave component of the Gaussian probe beam through the atomic medium. For the central plane-wave component, corresponding to $\Theta_x=\Theta_y=0$, the optical response of the medium is described by the Jones matrix
\begin{equation}
M_0=
\begin{pmatrix}
e^{i\phi_x(\theta)} & 0\\
0 & e^{i\phi_y}
\end{pmatrix},
\notag
\end{equation}
where $\phi_x(\theta)$ and $\phi_y$ are the generally complex phase shifts accumulated by the $x$- and $y$-polarized components, respectively.

Because a Gaussian beam comprises a continuous angular spectrum of plane-wave components, components propagating away from the central wave vector generally experience slightly different optical responses. An in-plane angular deviation $\Theta_x$ changes the angle between the propagation direction of the probe field and the quantization axis from $\theta$ to $\theta+\Theta_x$. Consequently,
\begin{equation}
\phi_x(\theta+\Theta_x)
\simeq
\phi_x(\theta)
+
\left.
\frac{\partial\phi_x}{\partial\theta}
\right|_{\theta}
\Theta_x,\notag
\end{equation}
to first order in $\Theta_x$.

By contrast, an out-of-plane angular deviation $\Theta_y$ causes the corresponding plane-wave component to propagate in a slightly rotated plane. This plane is related to that of the central component by a small azimuthal rotation about the quantization axis,
\begin{equation}
\varphi\simeq\frac{\Theta_y}{\sin\theta},\notag
\end{equation}
which results in a mixing of the two local polarization eigenvectors~\cite{bliokh2016spin}. 
The Jones matrix for an arbitrary angular-spectrum component can therefore be written as
\begin{equation}
\begin{aligned}
\widetilde{M}&(\Theta_x,\Theta_y)
=
\begin{pmatrix}
1 & \phi_{\bf B}\\
-\phi_{\bf B} & 1
\end{pmatrix}^{-1}
\begin{pmatrix}
e^{i\phi_x(\theta+\Theta_x)} & 0\\
0 & e^{i\phi_y}
\end{pmatrix}
\begin{pmatrix}
1 & \phi_{\bf B}\\
-\phi_{\bf B} & 1
\end{pmatrix}
\\[6pt]
&\simeq
\begin{pmatrix}
e^{i\phi_x(\theta+\Theta_x)}
&
\phi_{\bf B}\!\left[
e^{i\phi_x(\theta+\Theta_x)}-e^{i\phi_y}
\right]
\\[6pt]
\phi_{\bf B}\!\left[
e^{i\phi_x(\theta+\Theta_x)}-e^{i\phi_y}
\right]
&
e^{i\phi_y}
\end{pmatrix},
\label{eq:Mtheta_rotation}
\end{aligned}
\end{equation}
where $\phi_{\bf B}=\Theta_y\cot\theta$ is the Berry geometric phase~\cite{Luo2011Layered,bliokh2013goos,Chen2020OptExpress}, and terms of order $\phi_{\bf B}^2$ have been neglected. Substituting $\phi_x(\theta+\Theta_x)$ into Eq.~(\ref{eq:Mtheta_rotation}) and retaining only terms linear in $\Theta_x$ and $\Theta_y$, while neglecting the mixed term $\Theta_x\Theta_y$, yields
Eq.~(\ref{eq:Mtheta_total}).


For an $x$-polarized incident probe beam, the angular-spectrum representation of the input field is
\begin{equation}
\widetilde{\mathbf{E}}_p(\Theta_x,\Theta_y)
=
\begin{pmatrix}
1\\
0
\end{pmatrix}
\widetilde{u}(\Theta_x,\Theta_y).\notag
\end{equation}
After propagation through the atomic medium, the transmitted angular spectrum is given, apart from an overall phase factor, by
\begin{small}
\begin{equation}
\widetilde{\mathbf{E}}'_p(\Theta_x,\Theta_y)
\propto
\begin{pmatrix}
1\\[8pt]
\Theta_y\left(1-e^{i\Phi}\right)\cot\theta
\end{pmatrix}
\widetilde{u}(\Theta_x,\Theta_y)
\exp\left(
i\Theta_x\frac{\partial\phi_x}{\partial\theta}
\right),
\label{eq:EP_prime_x}
\end{equation}
\end{small}
this transmitted field can be rewritten as Eq.~(\ref{eq:EP_x_prime}).

Introducing the circular-polarization basis
\begin{equation}
\hat{\mathbf e}_{\pm}
=
\frac{1}{\sqrt{2}}
\left(
\hat{\mathbf x}\pm i\hat{\mathbf y}
\right)\notag
\end{equation}
and using $\Theta_{x,y}=k_{x,y}/k$, Eq.~(\ref{eq:EP_prime_x}) can be recast, to first order in the angular deviations, as
\begin{equation}
\begin{aligned}
\widetilde{\mathbf E}'_p(\Theta_x,\Theta_y)
\propto{}&
\widetilde{u}(\Theta_x,\Theta_y)
\Bigl[
e^{-i(k_x\Delta_x^{\hat{\mathbf{x}}}+k_y\Delta_y^{\hat{\mathbf{x}}})}
\hat{\mathbf e}_{+}
\\
&\qquad+
e^{-i(k_x\Delta_x^{\hat{\mathbf{x}}}-k_y\Delta_y^{\hat{\mathbf{x}}})}
\hat{\mathbf e}_{-}
\Bigr].
\end{aligned}
\label{eq:angular_circular_shift_x}
\end{equation}
Here, the first-order approximation requires
\begin{equation}
|\Delta_x^{\hat{\mathbf{x}}}|,\ |\Delta_y^{\hat{\mathbf{x}}}|\ll w_0.\notag
\end{equation}
Equation~(\ref{eq:angular_circular_shift_x}) shows that the two circularly polarized components acquire opposite transverse momentum-dependent phases, while sharing the same longitudinal phase shift. 
Finally, performing the inverse Fourier transform,
\begin{equation}
\mathbf E'_p(x,y)
\propto
\iint
\widetilde{\mathbf E}'_p(\Theta_x,\Theta_y)
\,d\Theta_x\,d\Theta_y,\notag
\end{equation}
yields the transmitted field in real space, as given in Eq.~(\ref{eq:realspace_x}).

Similarly, for a $y$-polarized incident probe beam, the input angular spectrum is given by
\begin{equation}
\widetilde{\mathbf E}_p(\Theta_x,\Theta_y)
=
\begin{pmatrix}
0\\
1
\end{pmatrix}
\widetilde{u}(\Theta_x,\Theta_y).\notag
\end{equation}
After propagation through the atomic medium, the transmitted angular spectrum becomes
\begin{equation}
\widetilde{\mathbf E}'_p(\Theta_x,\Theta_y)
\propto
\begin{pmatrix}
-\Theta_y\left(1-e^{-i\Phi}\right)\cot\theta\\[6pt]
1
\end{pmatrix}
\widetilde{u}(\Theta_x,\Theta_y).
\label{eq:EP_prime_y_vector}
\end{equation}
Using $\Theta_y=k_y/k$, Eq.~(\ref{eq:EP_prime_y_vector}) can be rewritten as
\begin{equation}
\widetilde{\mathbf E}'_p(\Theta_x,\Theta_y)
\propto
\widetilde{u}(\Theta_x,\Theta_y)
\left(
\hat{\mathbf{y}}
-k_y\Delta_y^{\hat{\mathbf{y}}}\hat{\mathbf{x}}
\right),
\label{eq:EP_prime_y}
\end{equation}
where the complex transverse shift is defined as
\begin{equation}
\Delta_y^{\hat{\mathbf{y}}}
=
\frac{1}{k}
\left(1-e^{-i\Phi}\right)\cot\theta.
\label{eq:complex_shift_y}\notag
\end{equation}
Unlike the $x$-polarized incident field, the $y$-polarized field does not acquire a longitudinal shift because $\phi_y$ is independent of the angle $\theta$. Finally, performing the inverse Fourier transform,
yields the transmitted field in real space, as given in Eq.~(\ref{eq:realspace_y}).



\section{Explicit expression of the optical Bloch equation}\label{appendixA}
The optical Bloch equation for the density-matrix (DM) elements $\rho_{\alpha\beta}({\bf r},t)=\langle\hat {S}_{\alpha\beta}({\bf r},t)\rangle$ is given by
\begin{subequations}\label{eqn1}
\begin{align}
&i\frac{\partial} {\partial t}\rho_{11}+i\Gamma_{21}\rho_{11}-i\Gamma_{12}\rho_{22}-i\Gamma_{13}
\rho_{33}+\Omega_{{p}1}^{\ast}\rho_{31}\notag\\
&\qquad-\Omega_{{p}1}\rho_{13}=0,\\
&i\frac{\partial} {\partial t}\rho_{22}+i\Gamma_{12}\rho_{22}-i\Gamma_{21}\rho_{11}-i\Gamma_{23}
\rho_{33}+\Omega_{{p}2}^{\ast}\rho_{32}\notag\\
&\qquad-\Omega_{{p}2}\rho_{23}=0,\\
&i\frac {\partial} {\partial t}\rho_{33}+i\Gamma_3\rho_{33}-i\Gamma_{34}\rho_{44}-\Omega_{{p}1}^{\ast}\rho_{31}
+\Omega_{{p}1}\rho_{13}\notag\\
&\qquad
-\Omega_{{p}2}^{\ast}\rho_{32}+\Omega_{{p}2}\rho_{23}
+\Omega_{{c}}^{\ast}\rho_{43}-\Omega_{{c}}\rho_{34}=0,\\
&i\frac {\partial} {\partial t}\rho_{44}+i\Gamma_{34}\rho_{44}-\Omega_{{c}}^{\ast}\rho_{43}
+\Omega_{{c}}\rho_{34}=0,\\
&\left(i\frac {\partial} {\partial t}+d_{21}\right)\rho_{21}+\Omega_{{p}2}^{\ast}\rho_{31}-\Omega_{{p}1}
\rho_{23}=0,\\
&\left(i\frac {\partial} {\partial t}+d_{31}\right)\rho_{31}+\Omega_{{c}}^{\ast}\rho_{41}
+\Omega_{{p}1}\left(\rho_{11}-\rho_{33}\right)+\Omega_{{p}2}
\rho_{21}=0,\\
&\left(i\frac {\partial} {\partial t}+d_{32}\right)\rho_{32}+\Omega_{{c}}^{\ast}\rho_{42}
+\Omega_{{p}2}\left(\rho_{22}-\rho_{33}\right)
+\Omega_{{p}1}\rho_{12}=0,\\
&\left(i\frac {\partial }{\partial t}+d_{41}\right)\rho_{41}+\Omega_{{c}}\rho_{31}-\Omega_{{p}1}\rho _{43}\notag\\
&\qquad-\mathcal{N}_a\int d^3r' V\left(\mathbf {r}'-\mathbf {r}\right)\rho_{44,41}\left( \mathbf {r'},\mathbf {r},t \right)=0,\\
&\left(i\frac{\partial}{\partial t }+d_{42}\right)\rho_{42}+\Omega_{{c}}\rho_{32}-\Omega_{{p}2}
\rho_{43}\notag\\
&\qquad-\mathcal {N}_{a }\int d^3r'V\left(\mathbf{r}'-\mathbf{r}\right)
\rho_{44,42}\left(\mathbf{r}',\mathbf{r},t\right)=0,\\
&\left(i\frac{\partial}{\partial t}+d_{43}\right)\rho_{43}+\Omega_{{c}}\left(\rho_{33}-\rho_{44}\right)
-\Omega_{{p}1}^{\ast }\rho_{41}-\Omega_{{p}2}^{\ast}\rho_{42}\notag\\
&\qquad
-\mathcal{N}_{a}\int  d^3r' V\left(\mathbf{r}' -\mathbf{r}\right)\rho_{44,43}\left(\mathbf{r}',\mathbf{r},t\right)=0,
\end{align}
\end{subequations}
Here, $d_{\alpha\beta }=\Delta _{\alpha }-\Delta _{\beta}+i\gamma_{\alpha\beta }$, $\gamma _{\alpha \beta}=\left(\Gamma _{\alpha }+\Gamma_{\beta} \right)/2+\gamma _{\alpha\beta}^{\rm{dep}}$,
$\Gamma_{\beta}=\sum_{\alpha <\beta}\Gamma _{\alpha\beta}$, with $\Gamma_{\alpha\beta}$ the spontaneous emission decay rate from the state $|\beta\rangle$ to the state $|\alpha\rangle$, $\gamma_{\alpha\beta}^{\rm{dep}}$ is the dephasing rate reflecting the loss of phase coherence between
$|\alpha\rangle$ and $|\beta\rangle$. The above equations for one-body DM elements involve two-body DM elements
$\rho_{44,4\alpha} \left(\mathbf{r}',\mathbf{r}, t\right)=\langle\hat{S}_{44}\left(\mathbf {r}', t\right)\hat{S}_{4\alpha}( \mathbf{r},t)\rangle$ are two-body DM elements.

\section{Steady state solution of the Bloch equation for one-body DM elements}\label{appendixB}

The expressions of the nonlinear susceptibilities of the system can be obtained by solving the Bloch equations~(\ref{eqn1}) under a steady-state approximation (i.e., taking $\partial{\rho_{\alpha\beta}}/\partial t=0$). We assume that atoms are initially populated in $|1\rangle$ and $|2\rangle$. Since the two probe fields are much weaker than the control field, we can take $\Omega_{pj}\sim \epsilon$ as an expansion parameter and  the perturbation expansion
$\rho_{\alpha\alpha}=\rho_{\alpha\alpha}^
{(0)}+\epsilon\rho_{\alpha\alpha}^{(1)} +\epsilon^{2}\rho_{\alpha\alpha}^{(2)} +\cdots$ ($\alpha=1-4$),
$\rho_{\alpha\beta} =\epsilon\rho_{\alpha\beta}^{(1)}+\epsilon^{2}\rho _{\alpha\beta}^{(2)}+\cdots$ ($\alpha= 2,\,3,\,4$; $\beta=1,\,2,\,3$; $\beta<\alpha$). Substituting the above expansions into Eqs.~(\ref{eqn1}), we obtain a set of linear but inhomogeneous equations which can be solved order by order~\cite{Mu2021PRA,Qin2020PRA,PhysRevA.110.013703}.

\subsection{First-order solutions}

At the first order of the expansion, the solution of $\rho_{31}^{(1)}$, $\rho_{41}^{(1)}$, $\rho_{32}^{(1)}$, $\rho_{42}^{(1)}$ reads
\begin{align}\label{rho1}
&\rho_{\alpha j}^{(1)}={a}_{\alpha j}^{(1)}\Omega_{{pj}},~~(\alpha=3,\,4; j=1,\,2)
\end{align}
where ${a}_{31}^{(1)}=d_{41}/(2D_1)$, ${a}_{32}^{(1)}=d_{42}/(2D_2)$, ${a}_{41}^{(1)}=-\Omega_c/(2D_1)$, ${a}_{42}^{(1)}=-\Omega_c/(2D_2)$, and $D_j=|\Omega_{{c}}|^2-d_{4j}d_{3j}$.

\subsection{Second-order solution}
At the second order of the expansion, $ \rho_{11}^{(2)}$, $\rho_{22}^{(2)}$, $\rho_{33}^{(2)}$, $\rho_{44}^{(2)}$, and $ \rho_{43}^{(2)}$ satisfy the equations
\begin{align}\label{rho2}
&\left( \begin{matrix}
i\Gamma_{21}&-i\Gamma_{12}&-i\Gamma_{13}&0&0&0\\
0&0& i\Gamma_{3}& -i\Gamma_{34}& \Omega_c^*&-\Omega_c\\
0&0&0& i\Gamma_{34}&-\Omega_c^*&\Omega_c\\
1& 1& 1& 1& 0&0\\
0&0&\Omega_c&-\Omega_c&d_{43}&0\\
0&0&\Omega_c^*&-\Omega_c^*&0&d_{43}^*\end{matrix} \right)
\left( \begin{matrix} \rho_{11}^{(2)}\\ \rho_{22}^{(2)}\\\rho_{33}^{(2)}\\\rho_{44}^{(2)}\\  \rho_{43}^{(2)}\\\rho_{34}^{(2)}\end{matrix} \right)
\notag\\
&
=\left( \begin{matrix}
\Omega_{{p}1}\rho_{13}^{(1)}-\Omega _{{p}1}^{\ast}\rho_{31}^{(1)}\\
2i{\rm Im}\left[\Omega_{{p}1}^*\rho_{31}^{(1)}+\Omega_{{p2}}^*\rho_{32}^{(1)}\right]\\
0\\
0\\
\Omega_{p1}^*\rho_{41}^{(1)}+\Omega_{p2}^*\rho_{42}^{(1)}\\
\Omega_{p1}\rho_{14}^{(1)}+\Omega_{p2}\rho_{24}^{(1)}
\end{matrix} \right).
\end{align}
Their solution has the form 
\begin{align}
\rho_{\alpha\beta}^{(2)}=a_{\alpha\beta,1}^{(2)}|\Omega_{{p1}}|^{2}
+a_{\alpha\beta,2}^{(2)}|\Omega_{{p2}}|^{2}.
\end{align}
The expression of $\rho_{21}^{(2)}$ reads
\begin{align}\label{12}
\rho_{21}^{(2)}&=\frac{\Omega_{{p}1}\rho_{23}^{(1)}-\Omega_{{p}2}^{*}\rho_{31}^{(1)}}{d_{21}}\equiv a_{21}^{(2)}\Omega_{p1}\Omega_{p2}^*.
\end{align}

\subsection{Third-order solution}
The solutions of $\rho_{31}^{(3)}$ and $\rho_{32}^{(3)}$ read
\begin{subequations}
\begin{align}\label{rho3}
\rho_{31}^{(3)}=&a_{31,1}^{(3)}|\Omega_{{p1}}|^{2}\Omega_{{p1}}
+a_{31,2}^{(3)}|\Omega_{{p2}}|^{2}\Omega_{{p1}}\notag\\
&+\mathcal{N}_a\int d^3r' \mathrm{V}\left({\bf r}^{\prime}-{\bf r}\right) a_{31,3}^{(3)}|\Omega_{{p1}}({\bf r}^{\prime})|^{2}\Omega_{{p1}}({\bf r})\notag\\
&+\mathcal{N}_a\int d^3r' \mathrm{V}\left({\bf r}^{\prime}-{\bf r}\right) a_{31,4}^{(3)}|\Omega_{{p2}}({\bf r}^{\prime})|^{2}\Omega_{{p1}}({\bf r}),\\
\rho_{32}^{(3)}=&a_{32,1}^{(3)}|\Omega_{{p1}}|^{2}\Omega_{{p2}}
+a_{32,2}^{(3)}|\Omega_{{p2}}|^{2}\Omega_{{p2}}\notag\\
&+\mathcal{N}_a\int d^3r' \mathrm{V}\left({\bf r}^{\prime}-{\bf r}\right) a_{32,3}^{(3)}|\Omega_{{p1}}({\bf r}^{\prime})|^{2}\Omega_{{p2}}({\bf r})\notag\\
&+\mathcal{N}_a\int d^3r' \mathrm{V}\left({\bf r}^{\prime}-{\bf r}\right) a_{32,4}^{(3)}|\Omega_{{p2}}({\bf r}^{\prime})|^{2}\Omega_{{p2}}({\bf r}),
\end{align}
\end{subequations}
with the coefficients given by 
\begin{subequations}\label{rho41local}
\begin{align}
&a_{31,l}^{(3)}=\frac{\Omega_{c}^{\ast}a_{43,l}^{(2)}+d_{41}\left(a_{11,l}^{(2)}-a_{33,l}^{^{(2)}}+\delta_{2l}a_{21}^{(2)}\right)}{D_1},\\
&a_{32,l}^{(3)}=\frac{\Omega_{c}^{\ast}a_{43,l}^{(2)}+d_{42}\left(a_{22,l}^{(2)}-a_{33,l}^{^{(2)}}+\delta_{1l} a_{12}^{(2)}\right)}{D_2},\\
&a_{31,l+2}^{(3)}=\frac{\Omega_{c}^{\ast}a_{44,41,l}^{(3)}}{D_1},\label{422}\\
&a_{32,l+2}^{(3)}=\frac{\Omega_{{c}}^{\ast} a_{44,42,l}^{(3)}}{D_2},
\end{align}
\end{subequations}
where $l=1,\,2$.

\section{Steady state solution of Bloch equations for two-body DM elements}

Note that the solution $\rho_{31}^{(3)}$ and $\rho_{32}^{(3)}$  include the parts of local terms and nonlocal (integral) terms contributed the Rydberg-Rydberg interaction. To get the nonlocal terms we must solve the equations of motion for two-body DM elements $\rho_{\alpha\beta,\mu\nu}$. 

\subsection{Second-order solution}
At the second-order approximation, the two-body DM elements $\rho_{41,41}^{(2)}$ and $\rho_{42,42}^{(2)}$
satisfy the equation
{\small\begin{align}\label{rho4141}
&\left(\begin{matrix}{2 d_{4\alpha}-V} & {2 \Omega_{{c}}} & {0} \\ {\Omega_{{c}}^{*}} & {d_{3\alpha}+d_{4\alpha}}  & {\Omega_{{c}}}  \\ {0} & {2\Omega_{{c}}^{*}} & {2d_{3\alpha}}\end{matrix}\right)
\left(\begin{matrix}{\rho_{4\alpha,4\alpha}^{(2)}} \\ {\rho_{4\alpha,3\alpha}^{(2)}} \\ {\rho_{3\alpha,3\alpha}^{(2)}} \end{matrix}\right)\notag\\
 &\quad=\left(\begin{matrix}{0} \\ {-\Omega_{p\alpha}\rho_{4\alpha}^{(1)}/2} \\ -\Omega_{p\alpha}{\rho_{3\alpha}^{(1)}}\end{matrix}\right),
\end{align}}
where $\alpha=1,\,2$; $\rho_{42,41}^{(2)}$ satisfies the equation
\begin{align}
&\left(\begin{matrix}
M & {\Omega_{{c}}} & {\Omega_{{c}}} & {0} \\
{\Omega_{{c}}^{*}} & {d_{42}+d_{31}} & {0} & {\Omega_{{c}}}  \\
{\Omega_{{c}}^{*}} & {0}  & {d_{41}+d_{32}} & {\Omega_{{c}}}  \\
{0} & {\Omega_{c}} & {\Omega_{{c}}^{*}} &  {d_{32}+d_{31}} \end{matrix}\right)
\left(\begin{matrix}{\rho_{42,41}^{(2)}} \\ {\rho_{42,31}^{(2)}} \\ {\rho_{41,32}^{(2)}} \\ {\rho_{32,31}^{(2)}}\end{matrix}\right)\notag\\
&=\left(\begin{matrix}
{0} \\ {-\Omega_{{p1}}\rho_{42}^{(1)}/2}\\
{-\Omega_{{p2}}\rho_{41}^{(1)}/2} \\ {-\Omega_{{p1}}\rho_{32}^{(1)}/2-\Omega_{{p2}}\rho_{31}^{(1)}/2} \end{matrix}\right),
\end{align}
with $M=d_{42}+d_{41}-V\left({\bf r}^{\prime}- {\bf r}\right)$.

\begin{widetext}
\subsection{Third-order solution}
The third-order two-body DM elements can be obtained by solving the equations ($\alpha=1,\,2$)
{\small \begin{align}\label{third-order two-body}
&\left(\begin{matrix}
{d_{3\alpha}+i\Gamma_3} & {\Omega_{c}^*} & {-i\Gamma_{34}}&{\Omega_{c}^*} &{0}&{0}&{-\Omega_{c}}&{0}  \\
{\Omega_{{c}}} & {d_{3\alpha}+d_{43}}  & {-\Omega_{c}}& {0}&{\Omega_{c}^*}& {0}& {0}& {0}  \\
{0} & {-\Omega_{{c}}^{*}} & {d_{3\alpha}+i\Gamma_{34}}&{0}&{0}&{\Omega_{c}^*}&{\Omega_{c}}&{0}\\
{\Omega_{c}}& {0}& {0}&{i\Gamma_{3}+d_{4\alpha}}&{\Omega_{c}^*}&{-i\Gamma_{34}}& {0}& {-\Omega_{c}}\\
{0}&{\Omega_{c}}& {0}&{\Omega_{c}}&{d_{4\alpha}+d_{43}}-V&{-\Omega_{c}}& {0}& {0}\\
{0}& {0}&{\Omega_{c}}& {0}&{-\Omega_{c}^*}&{i\Gamma_{34}+d_{4\alpha}}-V&{0}&{\Omega_{c}}\\
{\Omega_{{c}}^{*}} &{0}&{-\Omega_{{c}}^{*}}&{0}&{0}&{0}&{d_{3\alpha}-d_{43}^*}&{\Omega_{c}^*}\\
{0}&{0}&{0}&{-\Omega_{{c}}^{*}} &{0}&{\Omega_{{c}}^{*}}&{\Omega_{{c}}}&{-d_{43}^*+d_{4\alpha}}
\end{matrix}\right)
\left(\begin{matrix}
{\rho_{3\alpha,33}^{(3)}} \\ {\rho_{3\alpha,43}^{(3)}} \\ {\rho_{3\alpha,44}^{(3)}} \\
{\rho_{4\alpha,33}^{(3)}} \\ {\rho_{4\alpha,43}^{(3)}} \\ {\rho_{4\alpha,44}^{(3)}} \\
{\rho_{3\alpha,34}^{(3)}} \\ {\rho_{4\alpha,34}^{(3)}} \\
\end{matrix}\right)
 \nonumber\\
& =\left(\begin{matrix}
{\Omega_{p1}^*\rho\rho_{31,3\alpha}^{(2)}-\Omega_{p1}d_{41}^*\rho_{3\alpha}^{(1)}/D_1^*+\Omega_{p2}^*\rho\rho_{32,3\alpha}^{(2)}-\Omega_{p2}d_{42}^*\rho_{3\alpha}^{(1)}/D_2^*-\Omega_{p1}\rho_{33}^{(2)}/2} \\ {\Omega_{p1}^*\rho\rho_{41,3\alpha}^{(2)}+\Omega_{p2}^*\rho\rho_{42,3\alpha}^{(2)}-\Omega_{p1}\rho_{43}^{(2)}/2} \\
{-\Omega_{p1}\rho_{44}^{(2)}/2}
\\
{\Omega_{p1}^*\rho\rho_{31,4\alpha}^{(2)}-\Omega_{p1}d_{41}^*\rho_{4\alpha}^{(1)}/D_1^*+\Omega_{p2}^*\rho\rho_{32,4\alpha}^{(2)}-\Omega_{p2}^*d_{42}^*\rho_{4\alpha}^{(1)}/D_2^*}
\\
{\Omega_{p1}^*\rho\rho_{41,4\alpha}^{(2)}+\Omega_{p2}^*\rho\rho_{42,4\alpha}^{(2)}}
\\
{0}
\\
{-\Omega_{p1}\rho_{14}^{(1)}\rho_{3\alpha}^{(1)}-\Omega_{p2}\rho_{24}^{(1)}\rho_{3\alpha}^{(1)}-\Omega_{p1}\rho_{34}^{(2)}/2}
\\
{-\Omega_{p1}\rho_{14}^{(1)}\rho_{4\alpha}^{(1)}-\Omega_{p2}\rho_{24}^{(1)}\rho_{4\alpha}^{(1)}}
\end{matrix}\right).
\end{align}
Through solving them, one can obtain the general expression of $a_{44,4j,l}^{(3)}\left(\bf{r}^{\prime}-\bf{r}\right)$, which reads
\begin{align}\label{eqn12}
&a_{44,4j,l}^{(3)}\left(\bf{r}^{\prime}-\bf{r}\right)=\frac{P_{jl0}+P_{jl1}V\left( \bf{r}^{\prime}-\bf{r}\right)+P_{jl2}V\left( \bf{r}^{\prime}-\bf{r}\right)^2}{Q_{jl0}\!+\!Q_{jl1}\left( \bf{r}^{\prime}-\bf{r}\right)\!+\!Q_{jl2}\left( \bf{r}^{\prime}-\bf{r}\right)^2\!+\!Q_{jl3}\left( \bf{r}^{\prime}-\bf{r}\right)^3},
\end{align}}\noindent
where $P_{jln}$  and $Q_{jln}$ ($j,\,l=1,\,2;\,n=0,\,1,\,2,\,3$) are functions of $\Gamma_{\alpha\beta}$, $\Delta_{\alpha}$, and $\Omega_{{c}}$. The explicit expressions of $P_{jln}$  and $Q_{jln}$  are lengthy thus omitted here.
\end{widetext}

\section{Optical susceptibility}\label{AP:CHI}

The electric polarization of the probe field in the atomic gas is given by
\begin{align}
\mathbf{P}_p
={}&
\mathcal{N}_a
\left\{
\left(
\mathbf{p}_{13}\rho_{31}
+
\mathbf{p}_{23}\rho_{32}
\right)
\exp\left[i({\bf k}_p z-\omega_p t)\right]
+\mathrm{c.c.}
\right\}\notag
\end{align}
which can be expressed by
\begin{align}
\mathbf{P}_p
=\varepsilon_0
\left(
{\bf e}_{-}\mathcal{E}_{p1}\chi_1
+
{\bf e}_{+}\mathcal{E}_{p2}\chi_2
\right)
\exp\left[i({\bf k}_p z-\omega_p t)\right]
+\mathrm{c.c.}.\notag
\end{align}
where
\begin{subequations}
\begin{align}
\chi_{1}=\frac{\mathcal{N}_a\left({\mathbf e}_{-}
\cdot\mathbf{p}_{13}\right)\rho_{31}}{
\varepsilon_0\mathcal{E}_{p1}},\notag\\
\chi_{2}=\frac{\mathcal{N}_a\left({\mathbf e}_{+}
\cdot\mathbf{p}_{23}\right)\rho_{32}}
{\varepsilon_0\mathcal{E}_{p2}}.\notag
\end{align}
\end{subequations}
Based on the perturbative solutions obtained above, 
with
\begin{equation}
\Omega_{pj}
=\frac{
\left({\mathbf e}_{-(+)}\cdot
\mathbf{p}_{j3}\right)
\mathcal{E}_{pj}
}{2\hbar},~~~(j=1,\,2)\notag
\end{equation}
then we obtain the
explicit expressions of the optical susceptibilities
\begin{equation}
\begin{aligned}
\chi_j
={}&
\chi_j^{\rm{(lin)}}
+
\chi_{j1}^{(3,\mathrm{loc})}
\left|\mathcal{E}_{p1}\right|^2
+
\chi_{j2}^{(3,\mathrm{loc})}
\left|\mathcal{E}_{p2}\right|^2
\\
&+
\chi_{j1}^{(3,\mathrm{nloc})}
\left|\mathcal{E}_{p1}\right|^2
+
\chi_{j2}^{(3,\mathrm{nloc})}
\left|\mathcal{E}_{p2}\right|^2,
\end{aligned}
\end{equation}
where
\begin{equation}
\chi_j^{\rm{(lin)}}
=
\frac{
\mathcal{N}_a
\left|\mathbf{p}_{j3}\right|^2
a_{3j}^{(1)}
}{
\varepsilon_0\hbar
}\notag
\end{equation}
are the first-order linear susceptibilities, whereas
\begin{subequations}
\begin{align}
&\chi_{jl}^{(3,\mathrm{loc})}
=\frac{\mathcal{N}_a\left|\mathbf{p}_{j3}\right|^4
a_{3j,l}^{(3)}}{\varepsilon_0\hbar^3},\notag\\
&\chi_{jl}^{(3,\mathrm{nloc})}=\frac{
\mathcal{N}_a^2\left|\mathbf{p}_{j3}\right|^4a_{3j,l+2}^{(3)}}{\varepsilon_0\hbar^3},\notag
\end{align}
\end{subequations}
are the local and nonlocal third-order nonlinear susceptibilities, respectively.


\begin{thebibliography}{66}%
\makeatletter
\providecommand \@ifxundefined [1]{%
 \@ifx{#1\undefined}
}%
\providecommand \@ifnum [1]{%
 \ifnum #1\expandafter \@firstoftwo
 \else \expandafter \@secondoftwo
 \fi
}%
\providecommand \@ifx [1]{%
 \ifx #1\expandafter \@firstoftwo
 \else \expandafter \@secondoftwo
 \fi
}%
\providecommand \natexlab [1]{#1}%
\providecommand \enquote  [1]{``#1''}%
\providecommand \bibnamefont  [1]{#1}%
\providecommand \bibfnamefont [1]{#1}%
\providecommand \citenamefont [1]{#1}%
\providecommand \href@noop [0]{\@secondoftwo}%
\providecommand \href [0]{\begingroup \@sanitize@url \@href}%
\providecommand \@href[1]{\@@startlink{#1}\@@href}%
\providecommand \@@href[1]{\endgroup#1\@@endlink}%
\providecommand \@sanitize@url [0]{\catcode `\\12\catcode `\$12\catcode
  `\&12\catcode `\#12\catcode `\^12\catcode `\_12\catcode `\%12\relax}%
\providecommand \@@startlink[1]{}%
\providecommand \@@endlink[0]{}%
\providecommand \url  [0]{\begingroup\@sanitize@url \@url }%
\providecommand \@url [1]{\endgroup\@href {#1}{\urlprefix }}%
\providecommand \urlprefix  [0]{URL }%
\providecommand \Eprint [0]{\href }%
\providecommand \doibase [0]{https://doi.org/}%
\providecommand \selectlanguage [0]{\@gobble}%
\providecommand \bibinfo  [0]{\@secondoftwo}%
\providecommand \bibfield  [0]{\@secondoftwo}%
\providecommand \translation [1]{[#1]}%
\providecommand \BibitemOpen [0]{}%
\providecommand \bibitemStop [0]{}%
\providecommand \bibitemNoStop [0]{.\EOS\space}%
\providecommand \EOS [0]{\spacefactor3000\relax}%
\providecommand \BibitemShut  [1]{\csname bibitem#1\endcsname}%
\let\auto@bib@innerbib\@empty
\bibitem [{\citenamefont {Bliokh}\ \emph {et~al.}(2015)\citenamefont {Bliokh},
  \citenamefont {Rodr{\'i}guez-Fortu{\~n}o}, \citenamefont {Nori},\ and\
  \citenamefont {Zayats}}]{Bliokh2015SpinOrbit}%
  \BibitemOpen
  \bibfield  {author} {\bibinfo {author} {\bibfnamefont {K.~Y.}\ \bibnamefont
  {Bliokh}}, \bibinfo {author} {\bibfnamefont {F.~J.}\ \bibnamefont
  {Rodr{\'i}guez-Fortu{\~n}o}}, \bibinfo {author} {\bibfnamefont
  {F.}~\bibnamefont {Nori}},\ and\ \bibinfo {author} {\bibfnamefont {A.~V.}\
  \bibnamefont {Zayats}},\ }\bibfield  {title} {\bibinfo {title} {Spin--orbit
  interactions of light},\ }\href {https://doi.org/10.1038/nphoton.2015.201}
  {\bibfield  {journal} {\bibinfo  {journal} {Nat. Photon.}\ }\textbf {\bibinfo
  {volume} {9}},\ \bibinfo {pages} {796} (\bibinfo {year} {2015})}\BibitemShut
  {NoStop}%
\bibitem [{\citenamefont {Marrucci}\ \emph {et~al.}(2006)\citenamefont
  {Marrucci}, \citenamefont {Manzo},\ and\ \citenamefont
  {Paparo}}]{marrucci2006optical}%
  \BibitemOpen
  \bibfield  {author} {\bibinfo {author} {\bibfnamefont {L.}~\bibnamefont
  {Marrucci}}, \bibinfo {author} {\bibfnamefont {C.}~\bibnamefont {Manzo}},\
  and\ \bibinfo {author} {\bibfnamefont {D.}~\bibnamefont {Paparo}},\
  }\bibfield  {title} {\bibinfo {title} {Optical spin-to-orbital angular
  momentum conversion in inhomogeneous anisotropic media},\ }\href
  {https://doi.org/10.1103/PhysRevLett.96.163905} {\bibfield  {journal}
  {\bibinfo  {journal} {Phys. Rev. Lett.}\ }\textbf {\bibinfo {volume} {96}},\
  \bibinfo {pages} {163905} (\bibinfo {year} {2006})}\BibitemShut {NoStop}%
\bibitem [{\citenamefont {Petersen}\ \emph {et~al.}(2014)\citenamefont
  {Petersen}, \citenamefont {Volz},\ and\ \citenamefont
  {Rauschenbeutel}}]{Petersen2014}%
  \BibitemOpen
  \bibfield  {author} {\bibinfo {author} {\bibfnamefont {J.}~\bibnamefont
  {Petersen}}, \bibinfo {author} {\bibfnamefont {J.}~\bibnamefont {Volz}},\
  and\ \bibinfo {author} {\bibfnamefont {A.}~\bibnamefont {Rauschenbeutel}},\
  }\bibfield  {title} {\bibinfo {title} {Chiral nanophotonic waveguide
  interface based on spin--orbit interaction of light},\ }\href
  {https://doi.org/10.1126/science.1257671} {\bibfield  {journal} {\bibinfo
  {journal} {Science}\ }\textbf {\bibinfo {volume} {346}},\ \bibinfo {pages}
  {67} (\bibinfo {year} {2014})}\BibitemShut {NoStop}%
\bibitem [{\citenamefont {Qin}\ \emph {et~al.}(2026)\citenamefont {Qin},
  \citenamefont {Hang}, \citenamefont {Li},\ and\ \citenamefont
  {Zhao}}]{Qin_2026}%
  \BibitemOpen
  \bibfield  {author} {\bibinfo {author} {\bibfnamefont {L.}~\bibnamefont
  {Qin}}, \bibinfo {author} {\bibfnamefont {C.}~\bibnamefont {Hang}}, \bibinfo
  {author} {\bibfnamefont {W.}~\bibnamefont {Li}},\ and\ \bibinfo {author}
  {\bibfnamefont {X.}~\bibnamefont {Zhao}},\ }\bibfield  {title} {\bibinfo
  {title} {Nonlinear spin–orbit interaction of light in a cold {R}ydberg atom
  medium},\ }\href {https://doi.org/10.1088/0256-307X/43/5/050303} {\bibfield
  {journal} {\bibinfo  {journal} {Chin. Phys. Lett.}\ }\textbf {\bibinfo
  {volume} {43}},\ \bibinfo {pages} {050303} (\bibinfo {year}
  {2026})}\BibitemShut {NoStop}%
\bibitem [{\citenamefont {Onoda}\ \emph {et~al.}(2004)\citenamefont {Onoda},
  \citenamefont {Murakami},\ and\ \citenamefont {Nagaosa}}]{Onoda2004Hall}%
  \BibitemOpen
  \bibfield  {author} {\bibinfo {author} {\bibfnamefont {M.}~\bibnamefont
  {Onoda}}, \bibinfo {author} {\bibfnamefont {S.}~\bibnamefont {Murakami}},\
  and\ \bibinfo {author} {\bibfnamefont {N.}~\bibnamefont {Nagaosa}},\
  }\bibfield  {title} {\bibinfo {title} {Hall effect of light},\ }\href
  {https://doi.org/10.1103/PhysRevLett.93.083901} {\bibfield  {journal}
  {\bibinfo  {journal} {Phys. Rev. Lett.}\ }\textbf {\bibinfo {volume} {93}},\
  \bibinfo {pages} {083901} (\bibinfo {year} {2004})}\BibitemShut {NoStop}%
\bibitem [{\citenamefont {Kim}\ \emph {et~al.}(2023)\citenamefont {Kim},
  \citenamefont {Yang}, \citenamefont {Lee}, \citenamefont {Kim}, \citenamefont
  {Kim},\ and\ \citenamefont {Rho}}]{Kim2023}%
  \BibitemOpen
  \bibfield  {author} {\bibinfo {author} {\bibfnamefont {M.}~\bibnamefont
  {Kim}}, \bibinfo {author} {\bibfnamefont {Y.}~\bibnamefont {Yang}}, \bibinfo
  {author} {\bibfnamefont {D.}~\bibnamefont {Lee}}, \bibinfo {author}
  {\bibfnamefont {Y.}~\bibnamefont {Kim}}, \bibinfo {author} {\bibfnamefont
  {H.}~\bibnamefont {Kim}},\ and\ \bibinfo {author} {\bibfnamefont
  {J.}~\bibnamefont {Rho}},\ }\bibfield  {title} {\bibinfo {title} {Spin {Hall}
  effect of light: From fundamentals to recent advancements},\ }\href
  {https://doi.org/10.1002/lpor.202200046} {\bibfield  {journal} {\bibinfo
  {journal} {Laser Photonics Rev.}\ }\textbf {\bibinfo {volume} {17}},\
  \bibinfo {pages} {2200046} (\bibinfo {year} {2023})}\BibitemShut {NoStop}%
\bibitem [{\citenamefont {Imbert}(1972)}]{Imbert1972}%
  \BibitemOpen
  \bibfield  {author} {\bibinfo {author} {\bibfnamefont {C.}~\bibnamefont
  {Imbert}},\ }\bibfield  {title} {\bibinfo {title} {Calculation and
  experimental proof of the transverse shift induced by total internal
  reflection of a circularly polarized light beam},\ }\href
  {https://doi.org/10.1103/PhysRevD.5.787} {\bibfield  {journal} {\bibinfo
  {journal} {Phys. Rev. D}\ }\textbf {\bibinfo {volume} {5}},\ \bibinfo {pages}
  {787} (\bibinfo {year} {1972})}\BibitemShut {NoStop}%
\bibitem [{\citenamefont {Korger}\ \emph {et~al.}(2014)\citenamefont {Korger},
  \citenamefont {Aiello}, \citenamefont {Chille}, \citenamefont {Banzer},
  \citenamefont {Wittmann}, \citenamefont {Lindlein}, \citenamefont
  {Marquardt},\ and\ \citenamefont {Leuchs}}]{korger2014observation}%
  \BibitemOpen
  \bibfield  {author} {\bibinfo {author} {\bibfnamefont {J.}~\bibnamefont
  {Korger}}, \bibinfo {author} {\bibfnamefont {A.}~\bibnamefont {Aiello}},
  \bibinfo {author} {\bibfnamefont {V.}~\bibnamefont {Chille}}, \bibinfo
  {author} {\bibfnamefont {P.}~\bibnamefont {Banzer}}, \bibinfo {author}
  {\bibfnamefont {C.}~\bibnamefont {Wittmann}}, \bibinfo {author}
  {\bibfnamefont {N.}~\bibnamefont {Lindlein}}, \bibinfo {author}
  {\bibfnamefont {C.}~\bibnamefont {Marquardt}},\ and\ \bibinfo {author}
  {\bibfnamefont {G.}~\bibnamefont {Leuchs}},\ }\bibfield  {title} {\bibinfo
  {title} {Observation of the geometric spin {Hall} effect of light},\ }\href
  {https://doi.org/10.1103/PhysRevLett.112.113902} {\bibfield  {journal}
  {\bibinfo  {journal} {Phys. Rev. Lett.}\ }\textbf {\bibinfo {volume} {112}},\
  \bibinfo {pages} {113902} (\bibinfo {year} {2014})}\BibitemShut {NoStop}%
\bibitem [{\citenamefont {Bliokh}\ and\ \citenamefont
  {Bliokh}(2004)}]{Bliokh2004Topological}%
  \BibitemOpen
  \bibfield  {author} {\bibinfo {author} {\bibfnamefont {K.~Y.}\ \bibnamefont
  {Bliokh}}\ and\ \bibinfo {author} {\bibfnamefont {Y.~P.}\ \bibnamefont
  {Bliokh}},\ }\bibfield  {title} {\bibinfo {title} {Topological spin transport
  of photons: the optical {Magnus} effect and {Berry} phase},\ }\href
  {https://doi.org/10.1016/j.physleta.2004.10.035} {\bibfield  {journal}
  {\bibinfo  {journal} {Phys. Lett. A}\ }\textbf {\bibinfo {volume} {333}},\
  \bibinfo {pages} {181} (\bibinfo {year} {2004})}\BibitemShut {NoStop}%
\bibitem [{\citenamefont {Bliokh}\ and\ \citenamefont
  {Bliokh}(2006)}]{Bliokh2006Conservation}%
  \BibitemOpen
  \bibfield  {author} {\bibinfo {author} {\bibfnamefont {K.~Y.}\ \bibnamefont
  {Bliokh}}\ and\ \bibinfo {author} {\bibfnamefont {Y.~P.}\ \bibnamefont
  {Bliokh}},\ }\bibfield  {title} {\bibinfo {title} {Conservation of angular
  momentum, transverse shift, and spin {Hall} effect in reflection and
  refraction of an electromagnetic wave packet},\ }\href
  {https://doi.org/10.1103/PhysRevLett.96.073903} {\bibfield  {journal}
  {\bibinfo  {journal} {Phys. Rev. Lett.}\ }\textbf {\bibinfo {volume} {96}},\
  \bibinfo {pages} {073903} (\bibinfo {year} {2006})}\BibitemShut {NoStop}%
\bibitem [{\citenamefont {Ling}\ \emph {et~al.}(2017)\citenamefont {Ling},
  \citenamefont {Zhou}, \citenamefont {Huang}, \citenamefont {Liu},
  \citenamefont {Qiu}, \citenamefont {Luo},\ and\ \citenamefont
  {Wen}}]{ling2017recent}%
  \BibitemOpen
  \bibfield  {author} {\bibinfo {author} {\bibfnamefont {X.}~\bibnamefont
  {Ling}}, \bibinfo {author} {\bibfnamefont {X.}~\bibnamefont {Zhou}}, \bibinfo
  {author} {\bibfnamefont {K.}~\bibnamefont {Huang}}, \bibinfo {author}
  {\bibfnamefont {Y.}~\bibnamefont {Liu}}, \bibinfo {author} {\bibfnamefont
  {C.-W.}\ \bibnamefont {Qiu}}, \bibinfo {author} {\bibfnamefont
  {H.}~\bibnamefont {Luo}},\ and\ \bibinfo {author} {\bibfnamefont
  {S.}~\bibnamefont {Wen}},\ }\bibfield  {title} {\bibinfo {title} {Recent
  advances in the spin {Hall} effect of light},\ }\href
  {https://doi.org/10.1088/1361-6633/aa5397} {\bibfield  {journal} {\bibinfo
  {journal} {Rep. Prog. Phys.}\ }\textbf {\bibinfo {volume} {80}},\ \bibinfo
  {pages} {066401} (\bibinfo {year} {2017})}\BibitemShut {NoStop}%
\bibitem [{\citenamefont {Sheng}\ \emph {et~al.}(2023)\citenamefont {Sheng},
  \citenamefont {Chen}, \citenamefont {Yuan}, \citenamefont {Liu},
  \citenamefont {Zhang}, \citenamefont {Jing}, \citenamefont {Kuang},\ and\
  \citenamefont {Zhou}}]{Sheng2023PSHE}%
  \BibitemOpen
  \bibfield  {author} {\bibinfo {author} {\bibfnamefont {L.}~\bibnamefont
  {Sheng}}, \bibinfo {author} {\bibfnamefont {Y.}~\bibnamefont {Chen}},
  \bibinfo {author} {\bibfnamefont {S.}~\bibnamefont {Yuan}}, \bibinfo {author}
  {\bibfnamefont {X.}~\bibnamefont {Liu}}, \bibinfo {author} {\bibfnamefont
  {Z.}~\bibnamefont {Zhang}}, \bibinfo {author} {\bibfnamefont
  {H.}~\bibnamefont {Jing}}, \bibinfo {author} {\bibfnamefont {L.-M.}\
  \bibnamefont {Kuang}},\ and\ \bibinfo {author} {\bibfnamefont
  {X.}~\bibnamefont {Zhou}},\ }\bibfield  {title} {\bibinfo {title} {Photonic
  spin {Hall} effect: Physics, manipulations, and applications},\ }\href
  {https://doi.org/10.1016/j.pquantelec.2023.100484} {\bibfield  {journal}
  {\bibinfo  {journal} {Prog. Quantum Electron.}\ }\textbf {\bibinfo {volume}
  {91--92}},\ \bibinfo {pages} {100484} (\bibinfo {year} {2023})}\BibitemShut
  {NoStop}%
\bibitem [{\citenamefont {Zhou}\ \emph
  {et~al.}(2012{\natexlab{a}})\citenamefont {Zhou}, \citenamefont {Ling},
  \citenamefont {Luo},\ and\ \citenamefont {Wen}}]{Zhou2012Graphene}%
  \BibitemOpen
  \bibfield  {author} {\bibinfo {author} {\bibfnamefont {X.}~\bibnamefont
  {Zhou}}, \bibinfo {author} {\bibfnamefont {X.}~\bibnamefont {Ling}}, \bibinfo
  {author} {\bibfnamefont {H.}~\bibnamefont {Luo}},\ and\ \bibinfo {author}
  {\bibfnamefont {S.}~\bibnamefont {Wen}},\ }\bibfield  {title} {\bibinfo
  {title} {Identifying graphene layers via spin {Hall} effect of light},\
  }\href {https://doi.org/10.1063/1.4772502} {\bibfield  {journal} {\bibinfo
  {journal} {Appl. Phys. Lett.}\ }\textbf {\bibinfo {volume} {101}},\ \bibinfo
  {pages} {251602} (\bibinfo {year} {2012}{\natexlab{a}})}\BibitemShut
  {NoStop}%
\bibitem [{\citenamefont {Zhou}\ \emph
  {et~al.}(2012{\natexlab{b}})\citenamefont {Zhou}, \citenamefont {Xiao},
  \citenamefont {Luo},\ and\ \citenamefont {Wen}}]{Zhou2012Nanometal}%
  \BibitemOpen
  \bibfield  {author} {\bibinfo {author} {\bibfnamefont {X.}~\bibnamefont
  {Zhou}}, \bibinfo {author} {\bibfnamefont {Z.}~\bibnamefont {Xiao}}, \bibinfo
  {author} {\bibfnamefont {H.}~\bibnamefont {Luo}},\ and\ \bibinfo {author}
  {\bibfnamefont {S.}~\bibnamefont {Wen}},\ }\bibfield  {title} {\bibinfo
  {title} {Experimental observation of the spin {Hall} effect of light on a
  nanometal film via weak measurements},\ }\href
  {https://doi.org/10.1103/PhysRevA.85.043809} {\bibfield  {journal} {\bibinfo
  {journal} {Phys. Rev. A}\ }\textbf {\bibinfo {volume} {85}},\ \bibinfo
  {pages} {043809} (\bibinfo {year} {2012}{\natexlab{b}})}\BibitemShut
  {NoStop}%
\bibitem [{\citenamefont {Chen}\ \emph
  {et~al.}(2020{\natexlab{a}})\citenamefont {Chen}, \citenamefont {Ling},
  \citenamefont {Shu}, \citenamefont {Luo},\ and\ \citenamefont
  {Wen}}]{Chen2020}%
  \BibitemOpen
  \bibfield  {author} {\bibinfo {author} {\bibfnamefont {S.}~\bibnamefont
  {Chen}}, \bibinfo {author} {\bibfnamefont {X.}~\bibnamefont {Ling}}, \bibinfo
  {author} {\bibfnamefont {W.}~\bibnamefont {Shu}}, \bibinfo {author}
  {\bibfnamefont {H.}~\bibnamefont {Luo}},\ and\ \bibinfo {author}
  {\bibfnamefont {S.}~\bibnamefont {Wen}},\ }\bibfield  {title} {\bibinfo
  {title} {Precision measurement of the optical conductivity of atomically thin
  crystals via the photonic spin {Hall} effect},\ }\href
  {https://doi.org/10.1103/PhysRevApplied.13.014057} {\bibfield  {journal}
  {\bibinfo  {journal} {Phys. Rev. Applied}\ }\textbf {\bibinfo {volume}
  {13}},\ \bibinfo {pages} {014057} (\bibinfo {year}
  {2020}{\natexlab{a}})}\BibitemShut {NoStop}%
\bibitem [{\citenamefont {Wang}\ \emph
  {et~al.}(2020{\natexlab{a}})\citenamefont {Wang}, \citenamefont {Rong},
  \citenamefont {Maguid}, \citenamefont {Kleiner},\ and\ \citenamefont
  {Hasman}}]{WangB2020}%
  \BibitemOpen
  \bibfield  {author} {\bibinfo {author} {\bibfnamefont {B.}~\bibnamefont
  {Wang}}, \bibinfo {author} {\bibfnamefont {K.}~\bibnamefont {Rong}}, \bibinfo
  {author} {\bibfnamefont {E.}~\bibnamefont {Maguid}}, \bibinfo {author}
  {\bibfnamefont {V.}~\bibnamefont {Kleiner}},\ and\ \bibinfo {author}
  {\bibfnamefont {E.}~\bibnamefont {Hasman}},\ }\bibfield  {title} {\bibinfo
  {title} {Probing nanoscale fluctuation of ferromagnetic meta-atoms with a
  stochastic photonic spin {Hall} effect},\ }\href
  {https://doi.org/10.1038/s41565-020-0670-0} {\bibfield  {journal} {\bibinfo
  {journal} {Nat. Nanotechnol.}\ }\textbf {\bibinfo {volume} {15}},\ \bibinfo
  {pages} {450} (\bibinfo {year} {2020}{\natexlab{a}})}\BibitemShut {NoStop}%
\bibitem [{\citenamefont {Neugebauer}\ \emph {et~al.}(2019)\citenamefont
  {Neugebauer}, \citenamefont {Nechayev}, \citenamefont {Vorndran},
  \citenamefont {Leuchs},\ and\ \citenamefont {Banzer}}]{Neugebauer2019}%
  \BibitemOpen
  \bibfield  {author} {\bibinfo {author} {\bibfnamefont {M.}~\bibnamefont
  {Neugebauer}}, \bibinfo {author} {\bibfnamefont {S.}~\bibnamefont
  {Nechayev}}, \bibinfo {author} {\bibfnamefont {M.}~\bibnamefont {Vorndran}},
  \bibinfo {author} {\bibfnamefont {G.}~\bibnamefont {Leuchs}},\ and\ \bibinfo
  {author} {\bibfnamefont {P.}~\bibnamefont {Banzer}},\ }\bibfield  {title}
  {\bibinfo {title} {Weak measurement enhanced spin {Hall} effect of light for
  particle displacement sensing},\ }\href
  {https://doi.org/10.1021/acs.nanolett.8b04219} {\bibfield  {journal}
  {\bibinfo  {journal} {Nano Lett.}\ }\textbf {\bibinfo {volume} {19}},\
  \bibinfo {pages} {422} (\bibinfo {year} {2019})}\BibitemShut {NoStop}%
\bibitem [{\citenamefont {Wang}\ \emph
  {et~al.}(2020{\natexlab{b}})\citenamefont {Wang}, \citenamefont {Zhou},
  \citenamefont {Zeng}, \citenamefont {Chen}, \citenamefont {Ling},
  \citenamefont {Shu}, \citenamefont {Luo},\ and\ \citenamefont
  {Wen}}]{Wang2020APL}%
  \BibitemOpen
  \bibfield  {author} {\bibinfo {author} {\bibfnamefont {R.}~\bibnamefont
  {Wang}}, \bibinfo {author} {\bibfnamefont {J.}~\bibnamefont {Zhou}}, \bibinfo
  {author} {\bibfnamefont {K.}~\bibnamefont {Zeng}}, \bibinfo {author}
  {\bibfnamefont {S.}~\bibnamefont {Chen}}, \bibinfo {author} {\bibfnamefont
  {X.}~\bibnamefont {Ling}}, \bibinfo {author} {\bibfnamefont {W.}~\bibnamefont
  {Shu}}, \bibinfo {author} {\bibfnamefont {H.}~\bibnamefont {Luo}},\ and\
  \bibinfo {author} {\bibfnamefont {S.}~\bibnamefont {Wen}},\ }\bibfield
  {title} {\bibinfo {title} {Ultrasensitive and real-time detection of chemical
  reaction rate based on the photonic spin {Hall} effect},\ }\href
  {https://doi.org/10.1063/1.5131183} {\bibfield  {journal} {\bibinfo
  {journal} {APL Photon.}\ }\textbf {\bibinfo {volume} {5}},\ \bibinfo {pages}
  {016105} (\bibinfo {year} {2020}{\natexlab{b}})}\BibitemShut {NoStop}%
\bibitem [{\citenamefont {Zhu}\ \emph {et~al.}(2019)\citenamefont {Zhu},
  \citenamefont {Lou}, \citenamefont {Zhou}, \citenamefont {Zhang},
  \citenamefont {Huang}, \citenamefont {Li}, \citenamefont {Luo}, \citenamefont
  {Wen}, \citenamefont {Zhu}, \citenamefont {Gong}, \citenamefont {Qiu},\ and\
  \citenamefont {Ruan}}]{Zhu2019Differentiation}%
  \BibitemOpen
  \bibfield  {author} {\bibinfo {author} {\bibfnamefont {T.}~\bibnamefont
  {Zhu}}, \bibinfo {author} {\bibfnamefont {Y.}~\bibnamefont {Lou}}, \bibinfo
  {author} {\bibfnamefont {Y.}~\bibnamefont {Zhou}}, \bibinfo {author}
  {\bibfnamefont {J.}~\bibnamefont {Zhang}}, \bibinfo {author} {\bibfnamefont
  {J.}~\bibnamefont {Huang}}, \bibinfo {author} {\bibfnamefont
  {Y.}~\bibnamefont {Li}}, \bibinfo {author} {\bibfnamefont {H.}~\bibnamefont
  {Luo}}, \bibinfo {author} {\bibfnamefont {S.}~\bibnamefont {Wen}}, \bibinfo
  {author} {\bibfnamefont {S.}~\bibnamefont {Zhu}}, \bibinfo {author}
  {\bibfnamefont {Q.}~\bibnamefont {Gong}}, \bibinfo {author} {\bibfnamefont
  {M.}~\bibnamefont {Qiu}},\ and\ \bibinfo {author} {\bibfnamefont
  {Z.}~\bibnamefont {Ruan}},\ }\bibfield  {title} {\bibinfo {title}
  {Generalized spatial differentiation from the spin {Hall} effect of light and
  its application in image processing of edge detection},\ }\href
  {https://doi.org/10.1103/PhysRevApplied.11.034043} {\bibfield  {journal}
  {\bibinfo  {journal} {Phys. Rev. Applied}\ }\textbf {\bibinfo {volume}
  {11}},\ \bibinfo {pages} {034043} (\bibinfo {year} {2019})}\BibitemShut
  {NoStop}%
\bibitem [{\citenamefont {Tang}\ \emph {et~al.}(2024)\citenamefont {Tang},
  \citenamefont {Li}, \citenamefont {Kim}, \citenamefont {Xiao}, \citenamefont
  {Wu}, \citenamefont {Badloe}, \citenamefont {Li},\ and\ \citenamefont
  {Rho}}]{Tang2024}%
  \BibitemOpen
  \bibfield  {author} {\bibinfo {author} {\bibfnamefont {P.}~\bibnamefont
  {Tang}}, \bibinfo {author} {\bibfnamefont {X.}~\bibnamefont {Li}}, \bibinfo
  {author} {\bibfnamefont {Y.}~\bibnamefont {Kim}}, \bibinfo {author}
  {\bibfnamefont {L.}~\bibnamefont {Xiao}}, \bibinfo {author} {\bibfnamefont
  {H.}~\bibnamefont {Wu}}, \bibinfo {author} {\bibfnamefont {T.}~\bibnamefont
  {Badloe}}, \bibinfo {author} {\bibfnamefont {G.}~\bibnamefont {Li}},\ and\
  \bibinfo {author} {\bibfnamefont {J.}~\bibnamefont {Rho}},\ }\bibfield
  {title} {\bibinfo {title} {Spin {Hall} effect of light in a dichroic
  polarizer for multifunctional edge detection},\ }\href
  {https://doi.org/10.1021/acsphotonics.4c01058} {\bibfield  {journal}
  {\bibinfo  {journal} {ACS Photon.}\ }\textbf {\bibinfo {volume} {11}},\
  \bibinfo {pages} {4170} (\bibinfo {year} {2024})}\BibitemShut {NoStop}%
\bibitem [{\citenamefont {He}\ \emph {et~al.}(2025)\citenamefont {He},
  \citenamefont {Fang}, \citenamefont {Li}, \citenamefont {Yuan}, \citenamefont
  {Luo},\ and\ \citenamefont {Wang}}]{HeY2025}%
  \BibitemOpen
  \bibfield  {author} {\bibinfo {author} {\bibfnamefont {Y.}~\bibnamefont
  {He}}, \bibinfo {author} {\bibfnamefont {F.}~\bibnamefont {Fang}}, \bibinfo
  {author} {\bibfnamefont {C.}~\bibnamefont {Li}}, \bibinfo {author}
  {\bibfnamefont {W.}~\bibnamefont {Yuan}}, \bibinfo {author} {\bibfnamefont
  {H.}~\bibnamefont {Luo}},\ and\ \bibinfo {author} {\bibfnamefont
  {X.}~\bibnamefont {Wang}},\ }\bibfield  {title} {\bibinfo {title}
  {Full-dimensional photonic spin-{Hall} spatial differential imaging},\ }\href
  {https://doi.org/10.1021/acsphotonics.5c00567} {\bibfield  {journal}
  {\bibinfo  {journal} {ACS Photon.}\ }\textbf {\bibinfo {volume} {12}},\
  \bibinfo {pages} {2861} (\bibinfo {year} {2025})}\BibitemShut {NoStop}%
\bibitem [{\citenamefont {He}\ \emph {et~al.}(2020)\citenamefont {He},
  \citenamefont {Zhou}, \citenamefont {Chen}, \citenamefont {Shu},
  \citenamefont {Luo},\ and\ \citenamefont {Wen}}]{HeS2020}%
  \BibitemOpen
  \bibfield  {author} {\bibinfo {author} {\bibfnamefont {S.}~\bibnamefont
  {He}}, \bibinfo {author} {\bibfnamefont {J.}~\bibnamefont {Zhou}}, \bibinfo
  {author} {\bibfnamefont {S.}~\bibnamefont {Chen}}, \bibinfo {author}
  {\bibfnamefont {W.}~\bibnamefont {Shu}}, \bibinfo {author} {\bibfnamefont
  {H.}~\bibnamefont {Luo}},\ and\ \bibinfo {author} {\bibfnamefont
  {S.}~\bibnamefont {Wen}},\ }\bibfield  {title} {\bibinfo {title} {Spatial
  differential operation and edge detection based on the geometric spin {Hall}
  effect of light},\ }\href {https://doi.org/10.1364/OL.386224} {\bibfield
  {journal} {\bibinfo  {journal} {Opt. Lett.}\ }\textbf {\bibinfo {volume}
  {45}},\ \bibinfo {pages} {877} (\bibinfo {year} {2020})}\BibitemShut
  {NoStop}%
\bibitem [{\citenamefont {Yin}\ \emph {et~al.}(2013)\citenamefont {Yin},
  \citenamefont {Ye}, \citenamefont {Rho}, \citenamefont {Wang},\ and\
  \citenamefont {Zhang}}]{Yin2013Photonic}%
  \BibitemOpen
  \bibfield  {author} {\bibinfo {author} {\bibfnamefont {X.}~\bibnamefont
  {Yin}}, \bibinfo {author} {\bibfnamefont {Z.}~\bibnamefont {Ye}}, \bibinfo
  {author} {\bibfnamefont {J.}~\bibnamefont {Rho}}, \bibinfo {author}
  {\bibfnamefont {Y.}~\bibnamefont {Wang}},\ and\ \bibinfo {author}
  {\bibfnamefont {X.}~\bibnamefont {Zhang}},\ }\bibfield  {title} {\bibinfo
  {title} {Photonic spin {Hall} effect at metasurfaces},\ }\href
  {https://doi.org/10.1126/science.1231758} {\bibfield  {journal} {\bibinfo
  {journal} {Science}\ }\textbf {\bibinfo {volume} {339}},\ \bibinfo {pages}
  {1405} (\bibinfo {year} {2013})}\BibitemShut {NoStop}%
\bibitem [{\citenamefont {He}\ \emph {et~al.}(2021)\citenamefont {He},
  \citenamefont {Xu}, \citenamefont {Gao}, \citenamefont {Zhang},\ and\
  \citenamefont {Zhang}}]{HeA2021}%
  \BibitemOpen
  \bibfield  {author} {\bibinfo {author} {\bibfnamefont {A.}~\bibnamefont
  {He}}, \bibinfo {author} {\bibfnamefont {Y.}~\bibnamefont {Xu}}, \bibinfo
  {author} {\bibfnamefont {B.}~\bibnamefont {Gao}}, \bibinfo {author}
  {\bibfnamefont {T.}~\bibnamefont {Zhang}},\ and\ \bibinfo {author}
  {\bibfnamefont {J.}~\bibnamefont {Zhang}},\ }\bibfield  {title} {\bibinfo
  {title} {Subwavelength broadband photonic spin {Hall} devices via optical
  slot antennas},\ }\href {https://doi.org/10.1002/lpor.202000460} {\bibfield
  {journal} {\bibinfo  {journal} {Laser Photonics Rev.}\ }\textbf {\bibinfo
  {volume} {15}},\ \bibinfo {pages} {2000460} (\bibinfo {year}
  {2021})}\BibitemShut {NoStop}%
\bibitem [{\citenamefont {Kapitanova}\ \emph {et~al.}(2014)\citenamefont
  {Kapitanova}, \citenamefont {Ginzburg}, \citenamefont
  {Rodr{\'i}guez-Fortu{\~n}o}, \citenamefont {Filonov}, \citenamefont
  {Voroshilov}, \citenamefont {Belov}, \citenamefont {Poddubny}, \citenamefont
  {Kivshar}, \citenamefont {Wurtz},\ and\ \citenamefont
  {Zayats}}]{Kapitanova2014}%
  \BibitemOpen
  \bibfield  {author} {\bibinfo {author} {\bibfnamefont {P.~V.}\ \bibnamefont
  {Kapitanova}}, \bibinfo {author} {\bibfnamefont {P.}~\bibnamefont
  {Ginzburg}}, \bibinfo {author} {\bibfnamefont {F.~J.}\ \bibnamefont
  {Rodr{\'i}guez-Fortu{\~n}o}}, \bibinfo {author} {\bibfnamefont {D.~S.}\
  \bibnamefont {Filonov}}, \bibinfo {author} {\bibfnamefont {P.~M.}\
  \bibnamefont {Voroshilov}}, \bibinfo {author} {\bibfnamefont {P.~A.}\
  \bibnamefont {Belov}}, \bibinfo {author} {\bibfnamefont {A.~N.}\ \bibnamefont
  {Poddubny}}, \bibinfo {author} {\bibfnamefont {Y.~S.}\ \bibnamefont
  {Kivshar}}, \bibinfo {author} {\bibfnamefont {G.~A.}\ \bibnamefont {Wurtz}},\
  and\ \bibinfo {author} {\bibfnamefont {A.~V.}\ \bibnamefont {Zayats}},\
  }\bibfield  {title} {\bibinfo {title} {Photonic spin {Hall} effect in
  hyperbolic metamaterials for polarization-controlled routing of subwavelength
  modes},\ }\href {https://doi.org/10.1038/ncomms4226} {\bibfield  {journal}
  {\bibinfo  {journal} {Nat. Commun.}\ }\textbf {\bibinfo {volume} {5}},\
  \bibinfo {pages} {3226} (\bibinfo {year} {2014})}\BibitemShut {NoStop}%
\bibitem [{\citenamefont {Hosten}\ and\ \citenamefont
  {Kwiat}(2008)}]{Hosten2008}%
  \BibitemOpen
  \bibfield  {author} {\bibinfo {author} {\bibfnamefont {O.}~\bibnamefont
  {Hosten}}\ and\ \bibinfo {author} {\bibfnamefont {P.}~\bibnamefont {Kwiat}},\
  }\bibfield  {title} {\bibinfo {title} {Observation of the spin {Hall} effect
  of light via weak measurements},\ }\href
  {https://doi.org/10.1126/science.1152697} {\bibfield  {journal} {\bibinfo
  {journal} {Science}\ }\textbf {\bibinfo {volume} {319}},\ \bibinfo {pages}
  {787} (\bibinfo {year} {2008})}\BibitemShut {NoStop}%
\bibitem [{\citenamefont {Gorodetski}\ \emph {et~al.}(2012)\citenamefont
  {Gorodetski}, \citenamefont {Bliokh}, \citenamefont {Stein}, \citenamefont
  {Genet}, \citenamefont {Shitrit}, \citenamefont {Kleiner}, \citenamefont
  {Hasman},\ and\ \citenamefont {Ebbesen}}]{Gorodetski2012PRL}%
  \BibitemOpen
  \bibfield  {author} {\bibinfo {author} {\bibfnamefont {Y.}~\bibnamefont
  {Gorodetski}}, \bibinfo {author} {\bibfnamefont {K.~Y.}\ \bibnamefont
  {Bliokh}}, \bibinfo {author} {\bibfnamefont {B.}~\bibnamefont {Stein}},
  \bibinfo {author} {\bibfnamefont {C.}~\bibnamefont {Genet}}, \bibinfo
  {author} {\bibfnamefont {N.}~\bibnamefont {Shitrit}}, \bibinfo {author}
  {\bibfnamefont {V.}~\bibnamefont {Kleiner}}, \bibinfo {author} {\bibfnamefont
  {E.}~\bibnamefont {Hasman}},\ and\ \bibinfo {author} {\bibfnamefont {T.~W.}\
  \bibnamefont {Ebbesen}},\ }\bibfield  {title} {\bibinfo {title} {Weak
  measurements of light chirality with a plasmonic slit},\ }\href
  {https://doi.org/10.1103/PhysRevLett.109.013901} {\bibfield  {journal}
  {\bibinfo  {journal} {Phys. Rev. Lett.}\ }\textbf {\bibinfo {volume} {109}},\
  \bibinfo {pages} {013901} (\bibinfo {year} {2012})}\BibitemShut {NoStop}%
\bibitem [{\citenamefont {Luo}\ \emph {et~al.}(2011)\citenamefont {Luo},
  \citenamefont {Ling}, \citenamefont {Zhou}, \citenamefont {Shu},
  \citenamefont {Wen},\ and\ \citenamefont {Fan}}]{Luo2011Layered}%
  \BibitemOpen
  \bibfield  {author} {\bibinfo {author} {\bibfnamefont {H.}~\bibnamefont
  {Luo}}, \bibinfo {author} {\bibfnamefont {X.}~\bibnamefont {Ling}}, \bibinfo
  {author} {\bibfnamefont {X.}~\bibnamefont {Zhou}}, \bibinfo {author}
  {\bibfnamefont {W.}~\bibnamefont {Shu}}, \bibinfo {author} {\bibfnamefont
  {S.}~\bibnamefont {Wen}},\ and\ \bibinfo {author} {\bibfnamefont
  {D.}~\bibnamefont {Fan}},\ }\bibfield  {title} {\bibinfo {title} {Enhancing
  or suppressing the spin {Hall} effect of light in layered nanostructures},\
  }\href {https://doi.org/10.1103/PhysRevA.84.033801} {\bibfield  {journal}
  {\bibinfo  {journal} {Phys. Rev. A}\ }\textbf {\bibinfo {volume} {84}},\
  \bibinfo {pages} {033801} (\bibinfo {year} {2011})}\BibitemShut {NoStop}%
\bibitem [{\citenamefont {Slobozhanyuk}\ \emph {et~al.}(2016)\citenamefont
  {Slobozhanyuk}, \citenamefont {Poddubny}, \citenamefont {Sinev},
  \citenamefont {Samusev}, \citenamefont {Yu}, \citenamefont {Kuznetsov},
  \citenamefont {Miroshnichenko},\ and\ \citenamefont
  {Kivshar}}]{Slobozhanyuk2016Topological}%
  \BibitemOpen
  \bibfield  {author} {\bibinfo {author} {\bibfnamefont {A.~P.}\ \bibnamefont
  {Slobozhanyuk}}, \bibinfo {author} {\bibfnamefont {A.~N.}\ \bibnamefont
  {Poddubny}}, \bibinfo {author} {\bibfnamefont {I.~S.}\ \bibnamefont {Sinev}},
  \bibinfo {author} {\bibfnamefont {A.~K.}\ \bibnamefont {Samusev}}, \bibinfo
  {author} {\bibfnamefont {Y.~F.}\ \bibnamefont {Yu}}, \bibinfo {author}
  {\bibfnamefont {A.~I.}\ \bibnamefont {Kuznetsov}}, \bibinfo {author}
  {\bibfnamefont {A.~E.}\ \bibnamefont {Miroshnichenko}},\ and\ \bibinfo
  {author} {\bibfnamefont {Y.~S.}\ \bibnamefont {Kivshar}},\ }\bibfield
  {title} {\bibinfo {title} {Enhanced photonic spin {Hall} effect with
  subwavelength topological edge states},\ }\href
  {https://doi.org/10.1002/lpor.201600042} {\bibfield  {journal} {\bibinfo
  {journal} {Laser Photonics Rev.}\ }\textbf {\bibinfo {volume} {10}},\
  \bibinfo {pages} {656} (\bibinfo {year} {2016})}\BibitemShut {NoStop}%
\bibitem [{\citenamefont {Waseem}\ \emph {et~al.}(2024)\citenamefont {Waseem},
  \citenamefont {Shah},\ and\ \citenamefont {Gao}}]{Waseem2024}%
  \BibitemOpen
  \bibfield  {author} {\bibinfo {author} {\bibfnamefont {M.}~\bibnamefont
  {Waseem}}, \bibinfo {author} {\bibfnamefont {M.}~\bibnamefont {Shah}},\ and\
  \bibinfo {author} {\bibfnamefont {X.}~\bibnamefont {Gao}},\ }\bibfield
  {title} {\bibinfo {title} {Gain-assisted control of the photonic spin {Hall}
  effect},\ }\href {https://doi.org/10.1103/PhysRevA.110.033104} {\bibfield
  {journal} {\bibinfo  {journal} {Phys. Rev. A}\ }\textbf {\bibinfo {volume}
  {110}},\ \bibinfo {pages} {033104} (\bibinfo {year} {2024})}\BibitemShut
  {NoStop}%
\bibitem [{\citenamefont {Shah}\ \emph {et~al.}(2025)\citenamefont {Shah},
  \citenamefont {Qamar},\ and\ \citenamefont {Waseem}}]{Shah2025}%
  \BibitemOpen
  \bibfield  {author} {\bibinfo {author} {\bibfnamefont {M.}~\bibnamefont
  {Shah}}, \bibinfo {author} {\bibfnamefont {S.}~\bibnamefont {Qamar}},\ and\
  \bibinfo {author} {\bibfnamefont {M.}~\bibnamefont {Waseem}},\ }\bibfield
  {title} {\bibinfo {title} {Coherent control of photonic spin {Hall} effect in
  a cavity},\ }\href {https://doi.org/10.1364/OE.569158} {\bibfield  {journal}
  {\bibinfo  {journal} {Opt. Express}\ }\textbf {\bibinfo {volume} {33}},\
  \bibinfo {pages} {39359} (\bibinfo {year} {2025})}\BibitemShut {NoStop}%
\bibitem [{\citenamefont {Liu}\ \emph {et~al.}(2025)\citenamefont {Liu},
  \citenamefont {Abbas}, \citenamefont {Zhang},\ and\ \citenamefont
  {Lai}}]{Liu2025RydbergPSHE}%
  \BibitemOpen
  \bibfield  {author} {\bibinfo {author} {\bibfnamefont {W.}~\bibnamefont
  {Liu}}, \bibinfo {author} {\bibfnamefont {M.}~\bibnamefont {Abbas}}, \bibinfo
  {author} {\bibfnamefont {P.}~\bibnamefont {Zhang}},\ and\ \bibinfo {author}
  {\bibfnamefont {J.}~\bibnamefont {Lai}},\ }\bibfield  {title} {\bibinfo
  {title} {Nonlocal nonlinear control of photonic spin {Hall} effect in
  strongly interacting {Rydberg} media},\ }\href
  {https://doi.org/10.1088/1367-2630/ae1b35} {\bibfield  {journal} {\bibinfo
  {journal} {New J. Phys.}\ }\textbf {\bibinfo {volume} {27}},\ \bibinfo
  {pages} {113202} (\bibinfo {year} {2025})}\BibitemShut {NoStop}%
\bibitem [{\citenamefont {Patnaik}\ and\ \citenamefont
  {Agarwal}(2000)}]{Patnaik2000}%
  \BibitemOpen
  \bibfield  {author} {\bibinfo {author} {\bibfnamefont {A.~K.}\ \bibnamefont
  {Patnaik}}\ and\ \bibinfo {author} {\bibfnamefont {G.~S.}\ \bibnamefont
  {Agarwal}},\ }\bibfield  {title} {\bibinfo {title} {Laser field induced
  birefringence and enhancement of magneto-optical rotation},\ }\href
  {https://doi.org/10.1016/S0030-4018(99)00530-1} {\bibfield  {journal}
  {\bibinfo  {journal} {Opt. Commun.}\ }\textbf {\bibinfo {volume} {179}},\
  \bibinfo {pages} {97} (\bibinfo {year} {2000})}\BibitemShut {NoStop}%
\bibitem [{\citenamefont {Datsyuk}\ \emph {et~al.}(2008)\citenamefont
  {Datsyuk}, \citenamefont {Sokolov}, \citenamefont {Kupriyanov},\ and\
  \citenamefont {Havey}}]{Datsyuk2008}%
  \BibitemOpen
  \bibfield  {author} {\bibinfo {author} {\bibfnamefont {V.~M.}\ \bibnamefont
  {Datsyuk}}, \bibinfo {author} {\bibfnamefont {I.~M.}\ \bibnamefont
  {Sokolov}}, \bibinfo {author} {\bibfnamefont {D.~V.}\ \bibnamefont
  {Kupriyanov}},\ and\ \bibinfo {author} {\bibfnamefont {M.~D.}\ \bibnamefont
  {Havey}},\ }\bibfield  {title} {\bibinfo {title} {Electromagnetically induced
  optical anisotropy of an ultracold atomic medium},\ }\href
  {https://doi.org/10.1103/PhysRevA.77.033823} {\bibfield  {journal} {\bibinfo
  {journal} {Phys. Rev. A}\ }\textbf {\bibinfo {volume} {77}},\ \bibinfo
  {pages} {033823} (\bibinfo {year} {2008})}\BibitemShut {NoStop}%
\bibitem [{\citenamefont {Wielandy}\ and\ \citenamefont
  {Gaeta}(1998)}]{Wielandy1998}%
  \BibitemOpen
  \bibfield  {author} {\bibinfo {author} {\bibfnamefont {S.}~\bibnamefont
  {Wielandy}}\ and\ \bibinfo {author} {\bibfnamefont {A.~L.}\ \bibnamefont
  {Gaeta}},\ }\bibfield  {title} {\bibinfo {title} {Coherent control of the
  polarization of an optical field},\ }\href
  {https://doi.org/10.1103/PhysRevLett.81.3359} {\bibfield  {journal} {\bibinfo
   {journal} {Phys. Rev. Lett.}\ }\textbf {\bibinfo {volume} {81}},\ \bibinfo
  {pages} {3359} (\bibinfo {year} {1998})}\BibitemShut {NoStop}%
\bibitem [{\citenamefont {Boozer}\ \emph {et~al.}(2007)\citenamefont {Boozer},
  \citenamefont {Miller}, \citenamefont {Northup}, \citenamefont {Boca},\ and\
  \citenamefont {Kimble}}]{Boozer2007}%
  \BibitemOpen
  \bibfield  {author} {\bibinfo {author} {\bibfnamefont {A.~D.}\ \bibnamefont
  {Boozer}}, \bibinfo {author} {\bibfnamefont {R.}~\bibnamefont {Miller}},
  \bibinfo {author} {\bibfnamefont {T.~E.}\ \bibnamefont {Northup}}, \bibinfo
  {author} {\bibfnamefont {A.}~\bibnamefont {Boca}},\ and\ \bibinfo {author}
  {\bibfnamefont {H.~J.}\ \bibnamefont {Kimble}},\ }\bibfield  {title}
  {\bibinfo {title} {Optical pumping via incoherent {Raman} transitions},\
  }\href {https://doi.org/10.1103/PhysRevA.76.063401} {\bibfield  {journal}
  {\bibinfo  {journal} {Phys. Rev. A}\ }\textbf {\bibinfo {volume} {76}},\
  \bibinfo {pages} {063401} (\bibinfo {year} {2007})}\BibitemShut {NoStop}%
\bibitem [{\citenamefont {Chen}\ \emph {et~al.}(2002)\citenamefont {Chen},
  \citenamefont {Lin}, \citenamefont {Chen},\ and\ \citenamefont
  {Yu}}]{Chen2002Quantization}%
  \BibitemOpen
  \bibfield  {author} {\bibinfo {author} {\bibfnamefont {Y.-W.}\ \bibnamefont
  {Chen}}, \bibinfo {author} {\bibfnamefont {C.-W.}\ \bibnamefont {Lin}},
  \bibinfo {author} {\bibfnamefont {Y.-C.}\ \bibnamefont {Chen}},\ and\
  \bibinfo {author} {\bibfnamefont {I.~A.}\ \bibnamefont {Yu}},\ }\bibfield
  {title} {\bibinfo {title} {Quantization axes in coherent two-field
  spectroscopy},\ }\href {https://doi.org/10.1364/JOSAB.19.001917} {\bibfield
  {journal} {\bibinfo  {journal} {J. Opt. Soc. Am. B}\ }\textbf {\bibinfo
  {volume} {19}},\ \bibinfo {pages} {1917} (\bibinfo {year}
  {2002})}\BibitemShut {NoStop}%
\bibitem [{\citenamefont {Gerasimov}\ \emph {et~al.}(2011)\citenamefont
  {Gerasimov}, \citenamefont {Sokolov}, \citenamefont {Kupriyanov},
  \citenamefont {Olave},\ and\ \citenamefont {Havey}}]{Gerasimov2011}%
  \BibitemOpen
  \bibfield  {author} {\bibinfo {author} {\bibfnamefont {L.~V.}\ \bibnamefont
  {Gerasimov}}, \bibinfo {author} {\bibfnamefont {I.~M.}\ \bibnamefont
  {Sokolov}}, \bibinfo {author} {\bibfnamefont {D.~V.}\ \bibnamefont
  {Kupriyanov}}, \bibinfo {author} {\bibfnamefont {R.~G.}\ \bibnamefont
  {Olave}},\ and\ \bibinfo {author} {\bibfnamefont {M.~D.}\ \bibnamefont
  {Havey}},\ }\bibfield  {title} {\bibinfo {title} {Optical control of diffuse
  light storage in an ultracold atomic gas},\ }\href
  {https://doi.org/10.1364/JOSAB.28.001459} {\bibfield  {journal} {\bibinfo
  {journal} {J. Opt. Soc. Am. B}\ }\textbf {\bibinfo {volume} {28}},\ \bibinfo
  {pages} {1459} (\bibinfo {year} {2011})}\BibitemShut {NoStop}%
\bibitem [{\citenamefont {Xue}\ \emph {et~al.}(2019)\citenamefont {Xue},
  \citenamefont {Hao}, \citenamefont {Jiao}, \citenamefont {Han}, \citenamefont
  {Bai}, \citenamefont {Zhao},\ and\ \citenamefont {Raithel}}]{Xue2019Rydberg}%
  \BibitemOpen
  \bibfield  {author} {\bibinfo {author} {\bibfnamefont {Y.}~\bibnamefont
  {Xue}}, \bibinfo {author} {\bibfnamefont {L.}~\bibnamefont {Hao}}, \bibinfo
  {author} {\bibfnamefont {Y.}~\bibnamefont {Jiao}}, \bibinfo {author}
  {\bibfnamefont {X.}~\bibnamefont {Han}}, \bibinfo {author} {\bibfnamefont
  {S.}~\bibnamefont {Bai}}, \bibinfo {author} {\bibfnamefont {J.}~\bibnamefont
  {Zhao}},\ and\ \bibinfo {author} {\bibfnamefont {G.}~\bibnamefont
  {Raithel}},\ }\bibfield  {title} {\bibinfo {title} {{Rydberg
  Electromagnetically Induced Transparency in a Large Hilbert Space}},\ }\href
  {https://doi.org/10.1103/PhysRevA.99.053426} {\bibfield  {journal} {\bibinfo
  {journal} {Phys. Rev. A}\ }\textbf {\bibinfo {volume} {99}},\ \bibinfo
  {pages} {053426} (\bibinfo {year} {2019})}\BibitemShut {NoStop}%
\bibitem [{\citenamefont {Wu}\ \emph {et~al.}(2020)\citenamefont {Wu},
  \citenamefont {Zhang}, \citenamefont {Zhu},\ and\ \citenamefont
  {Agarwal}}]{Wu2020spin}%
  \BibitemOpen
  \bibfield  {author} {\bibinfo {author} {\bibfnamefont {J.}~\bibnamefont
  {Wu}}, \bibinfo {author} {\bibfnamefont {J.}~\bibnamefont {Zhang}}, \bibinfo
  {author} {\bibfnamefont {S.}~\bibnamefont {Zhu}},\ and\ \bibinfo {author}
  {\bibfnamefont {G.~S.}\ \bibnamefont {Agarwal}},\ }\bibfield  {title}
  {\bibinfo {title} {Spin-{Hall} effect of light and its enhancement in
  multilevel atomic system},\ }\href {https://doi.org/10.1364/OL.45.000149}
  {\bibfield  {journal} {\bibinfo  {journal} {Opt. Lett.}\ }\textbf {\bibinfo
  {volume} {45}},\ \bibinfo {pages} {149} (\bibinfo {year} {2020})}\BibitemShut
  {NoStop}%
\bibitem [{\citenamefont {Wu}\ \emph {et~al.}(2019)\citenamefont {Wu},
  \citenamefont {Zhang}, \citenamefont {Zhu},\ and\ \citenamefont
  {Agarwal}}]{Wu2019arxiv}%
  \BibitemOpen
  \bibfield  {author} {\bibinfo {author} {\bibfnamefont {J.}~\bibnamefont
  {Wu}}, \bibinfo {author} {\bibfnamefont {J.}~\bibnamefont {Zhang}}, \bibinfo
  {author} {\bibfnamefont {S.}~\bibnamefont {Zhu}},\ and\ \bibinfo {author}
  {\bibfnamefont {G.~S.}\ \bibnamefont {Agarwal}},\ }\bibfield  {title}
  {\bibinfo {title} {Controllable spin-{Hall} and related effects of light in
  an atomic medium via coupling fields},\ }\href
  {https://arxiv.org/abs/1901.00096} {\bibfield  {journal} {\bibinfo  {journal}
  {arXiv:1901.00096}\ } (\bibinfo {year} {2019})}\BibitemShut {NoStop}%
\bibitem [{\citenamefont {Cloutman}\ \emph {et~al.}(2026)\citenamefont
  {Cloutman}, \citenamefont {Chilcott}, \citenamefont {Elliott}, \citenamefont
  {Otto}, \citenamefont {Deb},\ and\ \citenamefont
  {Kj{\ae}rgaard}}]{Cloutman2026RydbergEIT}%
  \BibitemOpen
  \bibfield  {author} {\bibinfo {author} {\bibfnamefont {M.}~\bibnamefont
  {Cloutman}}, \bibinfo {author} {\bibfnamefont {M.}~\bibnamefont {Chilcott}},
  \bibinfo {author} {\bibfnamefont {A.}~\bibnamefont {Elliott}}, \bibinfo
  {author} {\bibfnamefont {J.~S.}\ \bibnamefont {Otto}}, \bibinfo {author}
  {\bibfnamefont {A.~B.}\ \bibnamefont {Deb}},\ and\ \bibinfo {author}
  {\bibfnamefont {N.}~\bibnamefont {Kj{\ae}rgaard}},\ }\bibfield  {title}
  {\bibinfo {title} {Electromagnetically induced transparency spectroscopy of
  {Rydberg} levels dressed by linearly polarized rf fields: Complementary
  angular response for two types of transition ladders},\ }\href
  {https://doi.org/10.1103/gcms-2yhg} {\bibfield  {journal} {\bibinfo
  {journal} {Phys. Rev. Research}\ }\textbf {\bibinfo {volume} {8}},\ \bibinfo
  {pages} {033089} (\bibinfo {year} {2026})}\BibitemShut {NoStop}%
\bibitem [{\citenamefont {Abbas}\ \emph {et~al.}(2025)\citenamefont {Abbas},
  \citenamefont {Zhang},\ and\ \citenamefont {Hamedi}}]{Abbas2025Tripod}%
  \BibitemOpen
  \bibfield  {author} {\bibinfo {author} {\bibfnamefont {M.}~\bibnamefont
  {Abbas}}, \bibinfo {author} {\bibfnamefont {P.}~\bibnamefont {Zhang}},\ and\
  \bibinfo {author} {\bibfnamefont {H.~R.}\ \bibnamefont {Hamedi}},\ }\bibfield
   {title} {\bibinfo {title} {Tunable photonic spin {Hall} effect in a tripod
  atom--light configuration},\ }\href
  {https://doi.org/10.1103/PhysRevA.111.043708} {\bibfield  {journal} {\bibinfo
   {journal} {Phys. Rev. A}\ }\textbf {\bibinfo {volume} {111}},\ \bibinfo
  {pages} {043708} (\bibinfo {year} {2025})}\BibitemShut {NoStop}%
\bibitem [{\citenamefont {Khan}\ \emph {et~al.}(2025)\citenamefont {Khan},
  \citenamefont {Du}, \citenamefont {Gao}, \citenamefont {Shah}, \citenamefont
  {Zhou},\ and\ \citenamefont {Gao}}]{Khan2025EIT}%
  \BibitemOpen
  \bibfield  {author} {\bibinfo {author} {\bibfnamefont {A.}~\bibnamefont
  {Khan}}, \bibinfo {author} {\bibfnamefont {K.}~\bibnamefont {Du}}, \bibinfo
  {author} {\bibfnamefont {L.}~\bibnamefont {Gao}}, \bibinfo {author}
  {\bibfnamefont {M.}~\bibnamefont {Shah}}, \bibinfo {author} {\bibfnamefont
  {X.}~\bibnamefont {Zhou}},\ and\ \bibinfo {author} {\bibfnamefont
  {D.}~\bibnamefont {Gao}},\ }\bibfield  {title} {\bibinfo {title} {Loss-free
  enhancement of photonic spin {Hall} shift by electromagnetically induced
  transparency},\ }\href {https://doi.org/10.1103/1c51-d7sz} {\bibfield
  {journal} {\bibinfo  {journal} {Phys. Rev. A}\ }\textbf {\bibinfo {volume}
  {111}},\ \bibinfo {pages} {063523} (\bibinfo {year} {2025})}\BibitemShut
  {NoStop}%
\bibitem [{\citenamefont {Wan}\ and\ \citenamefont {Zubairy}(2020)}]{Wan2020}%
  \BibitemOpen
  \bibfield  {author} {\bibinfo {author} {\bibfnamefont {R.-G.}\ \bibnamefont
  {Wan}}\ and\ \bibinfo {author} {\bibfnamefont {M.~S.}\ \bibnamefont
  {Zubairy}},\ }\bibfield  {title} {\bibinfo {title} {Controlling photonic spin
  {Hall} effect based on tunable surface plasmon resonance with an {N}-type
  coherent medium},\ }\href {https://doi.org/10.1103/PhysRevA.101.033837}
  {\bibfield  {journal} {\bibinfo  {journal} {Phys. Rev. A}\ }\textbf {\bibinfo
  {volume} {101}},\ \bibinfo {pages} {033837} (\bibinfo {year}
  {2020})}\BibitemShut {NoStop}%
\bibitem [{\citenamefont {Boyd}(2008)}]{Boyd2008}%
  \BibitemOpen
  \bibfield  {author} {\bibinfo {author} {\bibfnamefont {R.~W.}\ \bibnamefont
  {Boyd}},\ }\href@noop {} {\emph {\bibinfo {title} {Nonlinear Optics}}},\
  \bibinfo {edition} {3rd}\ ed.\ (\bibinfo  {publisher} {Academic Press},\
  \bibinfo {address} {New York},\ \bibinfo {year} {2008})\BibitemShut {NoStop}%
\bibitem [{\citenamefont {Fleischhauer}\ \emph {et~al.}(2005)\citenamefont
  {Fleischhauer}, \citenamefont {Imamoglu},\ and\ \citenamefont
  {Marangos}}]{Fleischhauer2005}%
  \BibitemOpen
  \bibfield  {author} {\bibinfo {author} {\bibfnamefont {M.}~\bibnamefont
  {Fleischhauer}}, \bibinfo {author} {\bibfnamefont {A.}~\bibnamefont
  {Imamoglu}},\ and\ \bibinfo {author} {\bibfnamefont {J.~P.}\ \bibnamefont
  {Marangos}},\ }\bibfield  {title} {\bibinfo {title} {Electromagnetically
  induced transparency: Optics in coherent media},\ }\href
  {https://doi.org/10.1103/RevModPhys.77.633} {\bibfield  {journal} {\bibinfo
  {journal} {Rev. Mod. Phys.}\ }\textbf {\bibinfo {volume} {77}},\ \bibinfo
  {pages} {633} (\bibinfo {year} {2005})}\BibitemShut {NoStop}%
\bibitem [{\citenamefont {Saffman}\ \emph {et~al.}(2010)\citenamefont
  {Saffman}, \citenamefont {Walker},\ and\ \citenamefont
  {M{\o}lmer}}]{Saffman2010}%
  \BibitemOpen
  \bibfield  {author} {\bibinfo {author} {\bibfnamefont {M.}~\bibnamefont
  {Saffman}}, \bibinfo {author} {\bibfnamefont {T.~G.}\ \bibnamefont
  {Walker}},\ and\ \bibinfo {author} {\bibfnamefont {K.}~\bibnamefont
  {M{\o}lmer}},\ }\bibfield  {title} {\bibinfo {title} {Quantum information
  with {Rydberg} atoms},\ }\href {https://doi.org/10.1103/RevModPhys.82.2313}
  {\bibfield  {journal} {\bibinfo  {journal} {Rev. Mod. Phys.}\ }\textbf
  {\bibinfo {volume} {82}},\ \bibinfo {pages} {2313} (\bibinfo {year}
  {2010})}\BibitemShut {NoStop}%
\bibitem [{\citenamefont {Gallagher}(2008)}]{Gallagher2008}%
  \BibitemOpen
  \bibfield  {author} {\bibinfo {author} {\bibfnamefont {T.~F.}\ \bibnamefont
  {Gallagher}},\ }\href@noop {} {\emph {\bibinfo {title} {Rydberg Atoms}}}\
  (\bibinfo  {publisher} {Cambridge University Press},\ \bibinfo {address}
  {Cambridge},\ \bibinfo {year} {2008})\BibitemShut {NoStop}%
\bibitem [{\citenamefont {Zhang}\ \emph {et~al.}(2025)\citenamefont {Zhang},
  \citenamefont {Zhang}, \citenamefont {Han}, \citenamefont {Zhang},
  \citenamefont {Zhang}, \citenamefont {Wu}, \citenamefont {Sovkov},
  \citenamefont {Liu}, \citenamefont {Li}, \citenamefont {Zhang}, \citenamefont
  {Xiao}, \citenamefont {Jia}, \citenamefont {Li},\ and\ \citenamefont
  {Ma}}]{Zhang2025Microwave}%
  \BibitemOpen
  \bibfield  {author} {\bibinfo {author} {\bibfnamefont {Z.}~\bibnamefont
  {Zhang}}, \bibinfo {author} {\bibfnamefont {Z.}~\bibnamefont {Zhang}},
  \bibinfo {author} {\bibfnamefont {S.}~\bibnamefont {Han}}, \bibinfo {author}
  {\bibfnamefont {Y.}~\bibnamefont {Zhang}}, \bibinfo {author} {\bibfnamefont
  {G.}~\bibnamefont {Zhang}}, \bibinfo {author} {\bibfnamefont
  {J.}~\bibnamefont {Wu}}, \bibinfo {author} {\bibfnamefont {V.~B.}\
  \bibnamefont {Sovkov}}, \bibinfo {author} {\bibfnamefont {W.}~\bibnamefont
  {Liu}}, \bibinfo {author} {\bibfnamefont {Y.}~\bibnamefont {Li}}, \bibinfo
  {author} {\bibfnamefont {L.}~\bibnamefont {Zhang}}, \bibinfo {author}
  {\bibfnamefont {L.}~\bibnamefont {Xiao}}, \bibinfo {author} {\bibfnamefont
  {S.}~\bibnamefont {Jia}}, \bibinfo {author} {\bibfnamefont {W.}~\bibnamefont
  {Li}},\ and\ \bibinfo {author} {\bibfnamefont {J.}~\bibnamefont {Ma}},\
  }\bibfield  {title} {\bibinfo {title} {Microwave-coupled optical bistability
  in driven and interacting {Rydberg} gases},\ }\href
  {https://doi.org/10.1038/s41534-025-00997-z} {\bibfield  {journal} {\bibinfo
  {journal} {npj Quantum Inf.}\ }\textbf {\bibinfo {volume} {11}},\ \bibinfo
  {pages} {44} (\bibinfo {year} {2025})}\BibitemShut {NoStop}%
\bibitem [{\citenamefont {Bai}\ and\ \citenamefont {Huang}(2016)}]{Bai2016OE}%
  \BibitemOpen
  \bibfield  {author} {\bibinfo {author} {\bibfnamefont {Z.}~\bibnamefont
  {Bai}}\ and\ \bibinfo {author} {\bibfnamefont {G.}~\bibnamefont {Huang}},\
  }\bibfield  {title} {\bibinfo {title} {Enhanced third-order and fifth-order
  {Kerr} nonlinearities in a cold atomic system via {Rydberg}-{Rydberg}
  interaction},\ }\href {https://doi.org/10.1364/OE.24.004442} {\bibfield
  {journal} {\bibinfo  {journal} {Opt. Express}\ }\textbf {\bibinfo {volume}
  {24}},\ \bibinfo {pages} {4442} (\bibinfo {year} {2016})}\BibitemShut
  {NoStop}%
\bibitem [{\citenamefont {Pritchard}\ \emph {et~al.}(2013)\citenamefont
  {Pritchard}, \citenamefont {Weatherill},\ and\ \citenamefont
  {Adams}}]{Pritchard2013in}%
  \BibitemOpen
  \bibfield  {author} {\bibinfo {author} {\bibfnamefont {J.~D.}\ \bibnamefont
  {Pritchard}}, \bibinfo {author} {\bibfnamefont {K.~J.}\ \bibnamefont
  {Weatherill}},\ and\ \bibinfo {author} {\bibfnamefont {C.~S.}\ \bibnamefont
  {Adams}},\ }\bibfield  {title} {\bibinfo {title} {Nonlinear optics using cold
  {Rydberg} atoms},\ }in\ \href {https://doi.org/10.1142/9789814440400_0008}
  {\emph {\bibinfo {booktitle} {Annual Review of Cold Atoms and Molecules}}},\
  Vol.~\bibinfo {volume} {1}\ (\bibinfo  {publisher} {World Scientific},\
  \bibinfo {address} {Singapore},\ \bibinfo {year} {2013})\ pp.\ \bibinfo
  {pages} {301--350}\BibitemShut {NoStop}%
\bibitem [{\citenamefont {Zheng}\ \emph {et~al.}(2022)\citenamefont {Zheng},
  \citenamefont {Zhao}, \citenamefont {Zhang},\ and\ \citenamefont
  {Wu}}]{WuJH2022PRA}%
  \BibitemOpen
  \bibfield  {author} {\bibinfo {author} {\bibfnamefont {D.-D.}\ \bibnamefont
  {Zheng}}, \bibinfo {author} {\bibfnamefont {H.-M.}\ \bibnamefont {Zhao}},
  \bibinfo {author} {\bibfnamefont {X.-J.}\ \bibnamefont {Zhang}},\ and\
  \bibinfo {author} {\bibfnamefont {J.-H.}\ \bibnamefont {Wu}},\ }\bibfield
  {title} {\bibinfo {title} {Correlation-enhanced {G}oos-{H}{\"a}nchen shift in
  {Rydberg} atomic gases},\ }\href
  {https://doi.org/10.1103/PhysRevA.106.043119} {\bibfield  {journal} {\bibinfo
   {journal} {Phys. Rev. A}\ }\textbf {\bibinfo {volume} {106}},\ \bibinfo
  {pages} {043119} (\bibinfo {year} {2022})}\BibitemShut {NoStop}%
\bibitem [{\citenamefont {Bliokh}\ \emph {et~al.}(2016)\citenamefont {Bliokh},
  \citenamefont {Samlan}, \citenamefont {Prajapati}, \citenamefont {Puentes},
  \citenamefont {Viswanathan},\ and\ \citenamefont {Nori}}]{bliokh2016spin}%
  \BibitemOpen
  \bibfield  {author} {\bibinfo {author} {\bibfnamefont {K.~Y.}\ \bibnamefont
  {Bliokh}}, \bibinfo {author} {\bibfnamefont {C.~T.}\ \bibnamefont {Samlan}},
  \bibinfo {author} {\bibfnamefont {C.}~\bibnamefont {Prajapati}}, \bibinfo
  {author} {\bibfnamefont {G.}~\bibnamefont {Puentes}}, \bibinfo {author}
  {\bibfnamefont {N.~K.}\ \bibnamefont {Viswanathan}},\ and\ \bibinfo {author}
  {\bibfnamefont {F.}~\bibnamefont {Nori}},\ }\bibfield  {title} {\bibinfo
  {title} {Spin-{Hall} effect and circular birefringence of a uniaxial crystal
  plate},\ }\href {https://doi.org/10.1364/OPTICA.3.001039} {\bibfield
  {journal} {\bibinfo  {journal} {Optica}\ }\textbf {\bibinfo {volume} {3}},\
  \bibinfo {pages} {1039} (\bibinfo {year} {2016})}\BibitemShut {NoStop}%
\bibitem [{\citenamefont {Bliokh}\ \emph {et~al.}(2019)\citenamefont {Bliokh},
  \citenamefont {Prajapati}, \citenamefont {Samlan}, \citenamefont
  {Viswanathan},\ and\ \citenamefont {Nori}}]{Bliokh2019}%
  \BibitemOpen
  \bibfield  {author} {\bibinfo {author} {\bibfnamefont {K.~Y.}\ \bibnamefont
  {Bliokh}}, \bibinfo {author} {\bibfnamefont {C.}~\bibnamefont {Prajapati}},
  \bibinfo {author} {\bibfnamefont {C.~T.}\ \bibnamefont {Samlan}}, \bibinfo
  {author} {\bibfnamefont {N.~K.}\ \bibnamefont {Viswanathan}},\ and\ \bibinfo
  {author} {\bibfnamefont {F.}~\bibnamefont {Nori}},\ }\bibfield  {title}
  {\bibinfo {title} {Spin-{Hall} effect of light at a tilted polarizer},\
  }\href {https://doi.org/10.1364/OL.44.004781} {\bibfield  {journal} {\bibinfo
   {journal} {Opt. Lett.}\ }\textbf {\bibinfo {volume} {44}},\ \bibinfo {pages}
  {4781} (\bibinfo {year} {2019})}\BibitemShut {NoStop}%
\bibitem [{\citenamefont {Ziauddin}\ \emph {et~al.}(2010)\citenamefont
  {Ziauddin}, \citenamefont {Qamar},\ and\ \citenamefont
  {Zubairy}}]{Ziauddin2010}%
  \BibitemOpen
  \bibfield  {author} {\bibinfo {author} {\bibnamefont {Ziauddin}}, \bibinfo
  {author} {\bibfnamefont {S.}~\bibnamefont {Qamar}},\ and\ \bibinfo {author}
  {\bibfnamefont {M.~S.}\ \bibnamefont {Zubairy}},\ }\bibfield  {title}
  {\bibinfo {title} {Coherent control of the {Goos--H{\"a}nchen} shift},\
  }\href {https://doi.org/10.1103/PhysRevA.81.023821} {\bibfield  {journal}
  {\bibinfo  {journal} {Phys. Rev. A}\ }\textbf {\bibinfo {volume} {81}},\
  \bibinfo {pages} {023821} (\bibinfo {year} {2010})}\BibitemShut {NoStop}%
\bibitem [{\citenamefont {Ziauddin}\ and\ \citenamefont
  {Qamar}(2011)}]{Ziauddin2011}%
  \BibitemOpen
  \bibfield  {author} {\bibinfo {author} {\bibnamefont {Ziauddin}}\ and\
  \bibinfo {author} {\bibfnamefont {S.}~\bibnamefont {Qamar}},\ }\bibfield
  {title} {\bibinfo {title} {Gain-assisted control of the {Goos--H{\"a}nchen}
  shift},\ }\href {https://doi.org/10.1103/PhysRevA.84.053844} {\bibfield
  {journal} {\bibinfo  {journal} {Phys. Rev. A}\ }\textbf {\bibinfo {volume}
  {84}},\ \bibinfo {pages} {053844} (\bibinfo {year} {2011})}\BibitemShut
  {NoStop}%
\bibitem [{\citenamefont {Mu}\ \emph {et~al.}(2021)\citenamefont {Mu},
  \citenamefont {Qin}, \citenamefont {Shi},\ and\ \citenamefont
  {Huang}}]{Mu2021PRA}%
  \BibitemOpen
  \bibfield  {author} {\bibinfo {author} {\bibfnamefont {Y.}~\bibnamefont
  {Mu}}, \bibinfo {author} {\bibfnamefont {L.}~\bibnamefont {Qin}}, \bibinfo
  {author} {\bibfnamefont {Z.}~\bibnamefont {Shi}},\ and\ \bibinfo {author}
  {\bibfnamefont {G.}~\bibnamefont {Huang}},\ }\bibfield  {title} {\bibinfo
  {title} {Giant {Kerr} nonlinearities and magneto-optical rotations in a
  {Rydberg}-atom gas via double electromagnetically induced transparency},\
  }\href {https://doi.org/10.1103/PhysRevA.103.043709} {\bibfield  {journal}
  {\bibinfo  {journal} {Phys. Rev. A}\ }\textbf {\bibinfo {volume} {103}},\
  \bibinfo {pages} {043709} (\bibinfo {year} {2021})}\BibitemShut {NoStop}%
\bibitem [{\citenamefont {Hang}\ and\ \citenamefont
  {Huang}(2018)}]{Hang2018Parity}%
  \BibitemOpen
  \bibfield  {author} {\bibinfo {author} {\bibfnamefont {C.}~\bibnamefont
  {Hang}}\ and\ \bibinfo {author} {\bibfnamefont {G.}~\bibnamefont {Huang}},\
  }\bibfield  {title} {\bibinfo {title} {Parity-time symmetry along with
  nonlocal optical solitons and their active controls in a {Rydberg} atomic
  gas},\ }\href {https://doi.org/10.1103/PhysRevA.98.043840} {\bibfield
  {journal} {\bibinfo  {journal} {Phys. Rev. A}\ }\textbf {\bibinfo {volume}
  {98}},\ \bibinfo {pages} {043840} (\bibinfo {year} {2018})}\BibitemShut
  {NoStop}%
\bibitem [{\citenamefont {Newell}\ and\ \citenamefont
  {Moloney}(1990)}]{Newell1980}%
  \BibitemOpen
  \bibfield  {author} {\bibinfo {author} {\bibfnamefont {A.~C.}\ \bibnamefont
  {Newell}}\ and\ \bibinfo {author} {\bibfnamefont {J.~V.}\ \bibnamefont
  {Moloney}},\ }\href@noop {} {\emph {\bibinfo {title} {Nonlinear Optics}}}\
  (\bibinfo  {publisher} {Addison-Wesley, Reading, MA},\ \bibinfo {year}
  {1990})\BibitemShut {NoStop}%
\bibitem [{\citenamefont {Singer}\ \emph {et~al.}(2005)\citenamefont {Singer},
  \citenamefont {Stanojevic}, \citenamefont {Weidem{\"u}ller},\ and\
  \citenamefont {C{\^o}t{\'e}}}]{Singer2005JPhysB}%
  \BibitemOpen
  \bibfield  {author} {\bibinfo {author} {\bibfnamefont {K.}~\bibnamefont
  {Singer}}, \bibinfo {author} {\bibfnamefont {J.}~\bibnamefont {Stanojevic}},
  \bibinfo {author} {\bibfnamefont {M.}~\bibnamefont {Weidem{\"u}ller}},\ and\
  \bibinfo {author} {\bibfnamefont {R.}~\bibnamefont {C{\^o}t{\'e}}},\
  }\bibfield  {title} {\bibinfo {title} {Long-range interactions between alkali
  {Rydberg} atom pairs correlated to the ns--ns, np--np and nd--nd
  asymptotes},\ }\href {https://doi.org/10.1088/0953-4075/38/2/021} {\bibfield
  {journal} {\bibinfo  {journal} {J. Phys. B: At. Mol. Opt. Phys.}\ }\textbf
  {\bibinfo {volume} {38}},\ \bibinfo {pages} {S295} (\bibinfo {year}
  {2005})}\BibitemShut {NoStop}%
\bibitem [{\citenamefont {Dai}\ \emph {et~al.}(2020)\citenamefont {Dai},
  \citenamefont {Yuan}, \citenamefont {Yin}, \citenamefont {Cao},\ and\
  \citenamefont {Chen}}]{PhysRevLett.124.053902}%
  \BibitemOpen
  \bibfield  {author} {\bibinfo {author} {\bibfnamefont {H.}~\bibnamefont
  {Dai}}, \bibinfo {author} {\bibfnamefont {L.}~\bibnamefont {Yuan}}, \bibinfo
  {author} {\bibfnamefont {C.}~\bibnamefont {Yin}}, \bibinfo {author}
  {\bibfnamefont {Z.}~\bibnamefont {Cao}},\ and\ \bibinfo {author}
  {\bibfnamefont {X.}~\bibnamefont {Chen}},\ }\bibfield  {title} {\bibinfo
  {title} {Direct visualizing the spin {Hall} effect of light via
  ultrahigh-order modes},\ }\href
  {https://doi.org/10.1103/PhysRevLett.124.053902} {\bibfield  {journal}
  {\bibinfo  {journal} {Phys. Rev. Lett.}\ }\textbf {\bibinfo {volume} {124}},\
  \bibinfo {pages} {053902} (\bibinfo {year} {2020})}\BibitemShut {NoStop}%
\bibitem [{\citenamefont {Bliokh}\ and\ \citenamefont
  {Aiello}(2013)}]{bliokh2013goos}%
  \BibitemOpen
  \bibfield  {author} {\bibinfo {author} {\bibfnamefont {K.~Y.}\ \bibnamefont
  {Bliokh}}\ and\ \bibinfo {author} {\bibfnamefont {A.}~\bibnamefont
  {Aiello}},\ }\bibfield  {title} {\bibinfo {title} {{Goos--H{\"a}nchen and
  Imbert--Fedorov beam shifts: an overview}},\ }\href
  {https://doi.org/10.1088/2040-8978/15/1/014001} {\bibfield  {journal}
  {\bibinfo  {journal} {J. Opt.}\ }\textbf {\bibinfo {volume} {15}},\ \bibinfo
  {pages} {014001} (\bibinfo {year} {2013})}\BibitemShut {NoStop}%
\bibitem [{\citenamefont {Chen}\ \emph
  {et~al.}(2020{\natexlab{b}})\citenamefont {Chen}, \citenamefont {Zhang},
  \citenamefont {Zhang}, \citenamefont {Li}, \citenamefont {Yang},
  \citenamefont {Zhang}, \citenamefont {Xi},\ and\ \citenamefont
  {Tang}}]{Chen2020OptExpress}%
  \BibitemOpen
  \bibfield  {author} {\bibinfo {author} {\bibfnamefont {Z.}~\bibnamefont
  {Chen}}, \bibinfo {author} {\bibfnamefont {H.}~\bibnamefont {Zhang}},
  \bibinfo {author} {\bibfnamefont {X.}~\bibnamefont {Zhang}}, \bibinfo
  {author} {\bibfnamefont {H.}~\bibnamefont {Li}}, \bibinfo {author}
  {\bibfnamefont {J.}~\bibnamefont {Yang}}, \bibinfo {author} {\bibfnamefont
  {W.}~\bibnamefont {Zhang}}, \bibinfo {author} {\bibfnamefont
  {L.}~\bibnamefont {Xi}},\ and\ \bibinfo {author} {\bibfnamefont
  {X.}~\bibnamefont {Tang}},\ }\bibfield  {title} {\bibinfo {title} {Symmetric
  spin splitting of elliptically polarized vortex beams reflected at an
  air--gold interface via the pseudo-{Brewster} angle},\ }\href
  {https://doi.org/10.1364/OE.403916} {\bibfield  {journal} {\bibinfo
  {journal} {Opt. Express}\ }\textbf {\bibinfo {volume} {28}},\ \bibinfo
  {pages} {29529} (\bibinfo {year} {2020}{\natexlab{b}})}\BibitemShut {NoStop}%
\bibitem [{\citenamefont {Qin}\ \emph {et~al.}(2020)\citenamefont {Qin},
  \citenamefont {Hang},\ and\ \citenamefont {Huang}}]{Qin2020PRA}%
  \BibitemOpen
  \bibfield  {author} {\bibinfo {author} {\bibfnamefont {L.}~\bibnamefont
  {Qin}}, \bibinfo {author} {\bibfnamefont {C.}~\bibnamefont {Hang}},\ and\
  \bibinfo {author} {\bibfnamefont {G.}~\bibnamefont {Huang}},\ }\bibfield
  {title} {\bibinfo {title} {High-fidelity and controllable cloning of
  high-dimensional optical beams with a {Rydberg} atomic gas},\ }\href
  {https://doi.org/10.1103/PhysRevA.102.063707} {\bibfield  {journal} {\bibinfo
   {journal} {Phys. Rev. A}\ }\textbf {\bibinfo {volume} {102}},\ \bibinfo
  {pages} {063707} (\bibinfo {year} {2020})}\BibitemShut {NoStop}%
\bibitem [{\citenamefont {Qin}\ \emph {et~al.}(2024)\citenamefont {Qin},
  \citenamefont {Hang}, \citenamefont {Huang},\ and\ \citenamefont
  {Li}}]{PhysRevA.110.013703}%
  \BibitemOpen
  \bibfield  {author} {\bibinfo {author} {\bibfnamefont {L.}~\bibnamefont
  {Qin}}, \bibinfo {author} {\bibfnamefont {C.}~\bibnamefont {Hang}}, \bibinfo
  {author} {\bibfnamefont {G.}~\bibnamefont {Huang}},\ and\ \bibinfo {author}
  {\bibfnamefont {W.}~\bibnamefont {Li}},\ }\bibfield  {title} {\bibinfo
  {title} {Shock-wave generation and propagation in dissipative and nonlocal
  nonlinear {R}ydberg media},\ }\href
  {https://doi.org/10.1103/PhysRevA.110.013703} {\bibfield  {journal} {\bibinfo
   {journal} {Phys. Rev. A}\ }\textbf {\bibinfo {volume} {110}},\ \bibinfo
  {pages} {013703} (\bibinfo {year} {2024})}\BibitemShut {NoStop}%
\end{thebibliography}
\end{document}